\documentclass[fleqn,usenatbib]{mnras}
\usepackage[T1]{fontenc}
\usepackage{ae,aecompl}
\usepackage{booktabs}

\usepackage{amssymb}	
\bibpunct{(}{)}{;}{a}{}{,} 
\usepackage[table]{xcolor}
\usepackage{etex}	
\usepackage{float}
\usepackage{mathtools}
\usepackage{ulem}
\usepackage{footmisc}
\usepackage[labelfont=bf, labelsep=period]{caption}

\usepackage{multicol}
\usepackage{tabto}

\definecolor{ochre}{rgb}{0.8, 0.47, 0.13}
\definecolor{orangered}{rgb}{1.0, 0.27, 0.0}

\usepackage{natbib}
\usepackage[utf8]{inputenc}
\usepackage{amsmath,units}
\usepackage{booktabs}
\usepackage{makecell} 
\usepackage{color}
\usepackage{tabularx}

\usepackage{acronym}  

\makeatletter
\newcommand{\acposs}[1]{%
 \expandafter\ifx\csname AC@#1\endcsname\AC@used
   \acs{#1}'s%
 \else
   \aclu{#1}'s (\acs{#1})%
 \fi
}
\makeatother

\usepackage{xspace}
\usepackage{enumitem} 
\usepackage{calc}
\usepackage[caption=false]{subfig}
\usepackage{import}

\usepackage{cleveref}
\setlist[enumerate,1]{label=(\roman*),ref=(\roman*),labelwidth =\widthof{\bfseries9999}, leftmargin = !}
\setlist[enumerate,2]{label=(\alph*),ref=(\roman{enumi}-\alph*),labelwidth =\widthof{\bfseries9999}, leftmargin = !}
\setlist[enumerate,3]{label=(\Alph*),ref=(\roman{enumi}-\alph{enumii}-\Alph*),labelwidth =\widthof{\bfseries9999}, leftmargin = !}
\setlist[enumerate,4]{label=(\arabic*),ref=(\roman{enumi}-\alph{enumii}-\Alph{enumiii}-\arabic*),labelwidth =\widthof{\bfseries9999}, leftmargin = !}

\usepackage{todonotes}

\newcommand{\PerCubicGigaparsecPerYear}{\,Gpc$^{\minus{3}}$~yr$^{\minus{1}}$}
\newcommand{\CubicGigaparsec}{Gpc$^{3}$}

\newcommand{\monei}{\ensuremath{m_{1,\rm{i}}}\xspace}
\newcommand{\mtwoi}{\ensuremath{m_{2,\rm{i}}}\xspace}

\newcommand{\qi}{\ensuremath{q_{\rm{i}}}\xspace}

\newcommand{\ei}{\ensuremath{{e}_{\rm{i}}}\xspace}

\newcommand{\Msun}{\ensuremath{\,\rm{M}_{\odot}}\xspace}

\newcommand{\AU}{\ensuremath{\,\mathrm{AU}}\xspace}

\newcommand{\Kelvin}{\ensuremath{\,\mathrm{K}}\xspace}

\newcommand{\dMdt}{\ensuremath{\frac{dM}{dt}}\xspace}

\newcommand\COMPAS{{\sc{COMPAS}}\xspace}

\newcommand{\minus}{\scalebox{0.75}[1.0]{$-$}}
\newcommand*{\logten}{\mathop{\log_{10}}\xspace}
\newcommand\tenpow[1]{\ensuremath{{\times}10^{#1}}}
\newcommand{\Zsun}{\ensuremath{\,\rm{Z}_{\odot}}\xspace}
\newcommand{\tsup}{\textsuperscript}

\makeatletter
\def\hlinewd#1{%
\noalign{\ifnum0=`}\fi\hrule \@height #1 %
\futurelet\reserved@a\@xhline}
\makeatother

\usepackage{array}
\newcolumntype{E}[1]{>{\raggedright\let\newline\\\arraybackslash\hspace{0pt}}m{#1}}
\newcolumntype{F}[1]{>{\centering\let\newline\\\arraybackslash\hspace{0pt}}m{#1}}
\newcolumntype{G}[1]{>{\raggedleft\let\newline\\\arraybackslash\hspace{0pt}}m{#1}}

\usepackage{lastpage}

\acrodef{CHE}{chemically homogeneous evolution}
\acrodef{CH}{chemically homogeneous}
\acrodef{MESA}{Modules for Experiments in Astrophysics}
\acrodef{ZAMS}{zero-age main sequence}
\acrodef{MS}{main sequence}
\acrodef{DCO}{double compact object}
\acrodef{BBH}{binary black hole}
\acrodefplural{BBH}[BBHs]{binary black holes}
\acrodef{RLOF}{Roche-lobe overflow}
\acrodef{CE}{common envelope}
\acrodef{CEE}{common envelope event}
\acrodef{SN}{supernova}
\acrodefplural{SN}[SNe]{supernovae}
\acrodef{ECSN}{electron-capture supernova}
\acrodefplural{ECSN}[ECSNe]{electron-capture supernovae}
\acrodef{USSN}{ultra-stripped supernova}
\acrodefplural{USSN}[USSNe]{ultra-stripped supernovae}
\acrodef{CCSN}{core-collapse supernova}
\acrodefplural{CCSN}[CCSNe]{core-collapse supernovae}
\acrodef{COMPAS}{Compact Object Mergers: Population Astrophysics and Statistics}
\acrodef{BPS}{binary population synthesis} 
\acrodef{SSE}{single star evolution} 
\acrodef{BSE}{binary star evolution} 
\acrodef{IMF}{initial mass function }
\acrodef{HR}{Hertzsprung-Russell}
\acrodef{HRD}{Hertzsprung-Russell diagram}
\acrodef{HeMS}{naked helium main sequence star}
\acrodef{LBV}{luminous blue variable}
\acrodef{WR}{Wolf-Rayet}
\acrodef{aLIGO}{Advanced Laser Interferometer Gravitational-wave Observatory}
\acrodef{PISN}{pair-instability supernova}
\acrodefplural{PISN}[PISNe]{pair-instability supernovae}
\acrodef{PPISN}{pulsational pair-instability supernova}
\acrodefplural{PPISN}[PPISNe]{pulsational pair-instability supernovae}
\acrodef{SFR}{star formation rate}
\acrodef{MSSFR}{metallicity-specific star formation rate}

\title{Chemically Homogeneous Evolution: A rapid population synthesis approach}

\author[J.~Riley et al.]{
	   Jeff Riley$^{1,2}$,
	   Ilya Mandel$^{1,2,3,4}$, 
	   Pablo Marchant$^5$, Ellen Butler$^3$, \newauthor
	   Kaila Nathaniel$^6$, Coenraad Neijssel$^{3,1}$, Spencer Shortt$^7$, 	   Alejandro Vigna-G\'{o}mez$^8$
\\
$^{1}$ School of Physics and Astronomy, Monash University, Clayton, Victoria 3800, Australia\\
$^{2}$ ARC Centre of Excellence for Gravitational Wave Discovery -- OzGrav, Australia\\	   
$^{3}$ Birmingham Institute for Gravitational Wave Astronomy and School of Physics and Astronomy,\\ University of Birmingham,  B15 2TT, Birmingham, UK\\
$^{4}$ ARC Centre of Excellence for All Sky Astrophysics in 3 Dimensions -- ASTRO 3D, Australia\\
$^{5}$ Institute of Astronomy, KU Leuven, Celestijnenlaan 200D, 3001 Leuven\\
$^{6}$ Argelander-Institut f\"{u}r Astronomie, Universit\"{a}t Bonn, Auf dem H\"{u}gel 71, 53121 Bonn, Germany,\\
$^{7}$ Department of Mathematics, University of Colorado Boulder, Boulder, CO, USA\\
$^{8}$ DARK, Niels Bohr Institute, University of Copenhagen, Jagtvej 128, 2200, Copenhagen, Denmark\\
\\
}

\date{Accepted 2021 April 30. Received 2021 April 14; in original form 2020 September 29}
\pubyear{2021}

\begin{document}
\label{FirstPage}

\pagerange{\pageref{FirstPage}--\pageref{LastPage}}
\maketitle

\begin{abstract}
We explore \ac{CHE} as a formation channel for massive merging \acp{BBH}.  We develop methods to include \ac{CHE} in a rapid binary population synthesis code, \ac{COMPAS}, which combines realistic models of binary evolution with cosmological models of the star-formation history of the Universe.  For the first time, we simultaneously explore conventional isolated binary star evolution under the same set of assumptions. This approach allows us to constrain population properties and make simultaneous predictions about the gravitational-wave detection rates of \ac{BBH} mergers for the \ac{CHE} and conventional formation channels. The overall mass distribution of detectable \acp{BBH} is consistent with existing gravitational-wave observations.  
We find that the \ac{CHE} channel may yield up to $\sim 70\%$ of all gravitational-wave detections of \ac{BBH} mergers coming from isolated binary evolution.  
\end{abstract}


\section{Introduction}\label{sec:intro}

On September 14th, 2015 the first direct observation of gravitational waves was made by the \acf{aLIGO} \citep{Abbott_2016}. The detected signal, now known as GW150914, was also the first observation of two black holes merging, thus confirming the existence of binary stellar-mass black hole systems and providing evidence that they can merge within the current age of the Universe. Based on 10 \acf{BBH} detections during the first two observing runs of \ac{aLIGO} and advanced Virgo, \citet{BBH:O1O2} estimate a local \ac{BBH} merger rate of 25\,--\,109\PerCubicGigaparsecPerYear at 90\% confidence.

How the \ac{BBH} sources of these gravitational wave signals form remains an open question. To be the source of gravitational waves detected at \ac{aLIGO}, which is sensitive to signals with frequencies of tens to hundreds of Hz, compact objects orbiting each other must spiral in as they lose energy through the emission of gravitational waves. Orbital energy loss through gravitational wave emission is not efficient at wide separations, and the timescale for gravitational wave emission to drive a binary to merger scales as the fourth power of the orbital separation \citep{Peters_1964}. In order for two $30\  M_\odot$ black holes to merge within $\approx 14$ Gyr, the current age of the Universe, their initial separation must be below $\lesssim 50\ R_\odot$.   And therein lies a problem, as this is smaller than the radial extent reached by typical slowly rotating massive stars during their evolution.  The different astrophysical channels proposed for forming merging \acp{BBH} generally fall into two categories \citep[see, e.g.,][for reviews]{MandelFarmer:2018,Mapelli:2018}:
\begin{enumerate}[label=(\roman*), labelwidth =\widthof{\bfseries9999}, leftmargin = !]
  \item isolated binary evolution, in which two stars may interact through tides and mass transfer, but are dynamically decoupled from other stars \citep[e.g.,][]{Tutukov:1973,Tutukov_1993, vdH:1976}.
  \item dynamical formation, where dynamical interactions in a dense environment and/or a hierarchical triple system play a key role in forming and hardening a compact \ac{BBH} \citep{Sigurdsson_1993, Miller_2009, Ziosi:2014, Rodriguez_2015, Antonini_2016, Bartos:2016, Stone:2016}.
\end{enumerate}

A variant of the isolated binary evolution channel relies on rotationally-induced chemical mixing in massive stars to prevent the establishment of a strong chemical gradient \citep{Maeder_1987, Maeder_2000, Heger_2000}. As long as the star continues to rotate at a sufficiently high rate, it will remain quasi-chemically homogeneous \citep{Maeder_1987, Langer_1992}. Contrary to the core-envelope structure exhibited by conventional, more slowly rotating stars, and the characteristic expansion of the envelope as the core contracts, the radius of quasi-chemically homogeneous stars will shrink or remain constant as they become hotter and brighter \citep{Yoon_2006,Mandel_2016}. 

As we discuss below, previous work on the \acf{CHE} channel for \ac{BBH} formation \citep{Mandel_2016,Marchant_2016, deMink_2016, duBuisson_2020} explored this channel independently of the usual isolated binary evolution channel.  In this paper, we present our rapid population synthesis model for the \ac{CHE} of binary systems, allowing for a direct comparison of the rates and properties of \ac{CHE} and non-\ac{CHE} \acp{BBH} under the same set of assumptions.   Our \ac{CHE} model is implemented in the rapid binary population synthesis code \ac{COMPAS} \citep{COMPAS_Stevenson_2017, COMPAS_VignaGomez_2018}, with thresholds on \ac{CHE} evolution computed using rotating stellar models in the \ac{MESA} code \citep{Paxton_2015, Paxton_2013, Paxton_2011}.

The remainder of this paper is organised as follows. Section~\ref{sec:che} is a brief outline of \ac{CHE} and previous work on the formation of \acp{BBH} through this channel.  Section~\ref{sec:methods} presents a description of our \ac{CHE} model and the implementation of the model in \ac{COMPAS}.  We present our results in Section~\ref{sec:results}.  We provide some concluding remarks in Section~\ref{sec:conclusion}.

\section{Chemically Homogeneous Evolution}\label{sec:che}

Stars evolving on the \ac{MS} typically develop increasingly helium-rich cores and hydrogen-rich envelopes as radial mixing is inefficient.  However, \citet{vonZeipel:1924} showed that rotating stars cannot simultaneously be in hydrostatic and thermal equilibrium if the rotational velocity is a function of  radius only, which has been argued to result in meridional currents in the radiative layers of a rotating star \citep{Sweet_1950, Eddington_1929}.  In massive rapidly rotating stars in low-metallicity environments, these currents can mix material from the convective core throughout the radiative envelope, leading to chemically homogeneous evolution for rapidly rotating stars \citep{Maeder_1987}.

Due to strong chemical mixing, chemically homogeneous stars do not maintain a hydrogen-rich envelope~--~thus avoiding the dramatic expansion exhibited during the post-main sequence phase by non-chemically homogeneous stars. The radius of a chemically homogeneous star remains stable, or shrinks slowly, as the star becomes increasingly helium rich over the course of the main sequence, with the star contracting to a massive naked helium star post-main sequence. Chemically homogeneous components of a very close binary system can thus avoid overfilling their Roche lobes, mass transfer, and probable merger.  

\citet{deMink_2009} modelled the evolution of rotating massive stars using the hydrodynamic stellar evolution code described by \citet{Yoon_2006} and \citet{Petrovic_2005}, which includes the effects of rotation on the stellar structure and the transport of angular momentum via rotationally-induced hydrodynamic instabilities \citep{Heger_2000}. The binary models developed by \citet{deMink_2009} and \citet{Song_2016} show that constituent stars in very tight binary systems can achieve rotational frequencies sufficient to induce \ac{CHE}.  \citet{deMink_2009} proposed \ac{CHE} as a viable formation channel for high-mass black-hole X-ray binaries.  VFTS~352 \citep{Almeida_2015} and HD~5980 \citep{Koenigsberger_2014} are examples of observed binary systems thought to have undergone \ac{CHE} \citep{deMink_2016}.

\citet{Mandel_2016} and \citet{Marchant_2016} introduced and investigated \ac{CHE} as a channel for forming merging \acp{BBH}.  They concluded that for sufficiently high masses and sufficiently low metallicities, a narrow range of initial orbital periods (short enough to allow rapid rotation necessary for \ac{CHE}, but not so short that the binary would immediately merge) could allow this channel to produce merging \acp{BBH}.

\citet{Mandel_2016,deMink_2016} used approximate thresholds for \ac{CHE} based on the models of \citet{Yoon_2006} to investigate the rates and properties of \acp{BBH} formed through the \ac{CHE} channel.  They estimated a merger rate of $\sim 10$  Gpc$^{-3}$ yr$^{-1}$ in the local Universe for this channel, subject to a number of evolutionary uncertainties, which they explored in a population-synthesis-style study.

\citet{Marchant_2016} used the MESA code to conduct detailed simulations of the \ac{CHE} channel, which were followed until the \ac{BBH} stage. The simulations were conducted for close binaries with component masses above $\sim20$\,\Msun, and included the over-contact phase in a majority of \ac{CHE} \ac{BBH} progenitors. \citet{Marchant_2016} suggested that as long as material does not overflow the L2 point in over-contact binaries, co-rotation can be maintained, and a spiral-in due to viscous drag can be avoided. In this scenario, close binary systems typically enter the over-contact phase in the early stages of core hydrogen burning, and then equilibrate their masses through mass transfer between the constituent stars. \citet{duBuisson_2020} extended the results of the MESA simulations performed by \citet{Marchant_2016} and combined them with the cosmological simulations of the chemical and star-formation history on the universe by \citet{Taylor_2015}.  Their population synthesis study investigated the population properties, cosmological rates and \ac{aLIGO} detection rates of \acp{BBH}, including the dependence on the early-Universe \ac{SFR}, which they find to be mild for moderate variations in the high-redshift \ac{SFR}.

\section{Methods}\label{sec:methods}

In this section, we describe the implementation of \ac{CHE} within the \ac{COMPAS} rapid binary population synthesis code.  Using \ac{COMPAS} allows us to rapidly evolve a large synthetic population of binaries, which includes binaries whose component stars evolve conventionally (i.e. along a redwards track on the \ac{HRD}), and others whose components evolve via \ac{CHE} (i.e. along a bluewards track on the \ac{HRD}), thus providing data for both pathways that can be compared directly.  Below, we summarise the key physics implemented in \ac{COMPAS}, starting with our approximate model of quasi-chemically homogeneous evolution based on MESA experiments, as well as the choices made for the metallicity-specific star formation history.  

\subsection{Physics implemented in COMPAS}\label{sec:COMPASphysics}

The basics of stellar and binary evolution and \ac{BBH} population modelling in COMPAS are described by \citet{COMPAS_Stevenson_2017, COMPAS_VignaGomez_2018, Neijssel_2019}.  Here, we provide a brief summary and describe differences from previous \ac{COMPAS} studies.

\subsubsection{\ac{CHE} in COMPAS}\label{sec:COMPASphysics_CHE}

We used a set of MESA models of single stars with a fixed rotational frequency and no mass loss to determine the minimal angular frequency $\omega$ necessary for chemically homogeneous evolution as a function of mass and metallicity.  Our fits to these angular frequency thresholds are provided in Appendix~\ref{sec:che_thresholds}.  \ac{COMPAS} uses these fits to determine whether a star is evolving chemically homogeneously.

Stellar evolution in \ac{COMPAS} follows the analytical fits of \citet{Hurley_2000} to the stellar models from \citet{Pols_1998}. In order to address \ac{CHE}, we introduce a new \ac{CH} stellar type to the \citet{Hurley_2000} collection of stellar types.   In our simplified model, we neglect the very limited radial evolution of a \ac{CH} star and set its radius equal to the \ac{ZAMS} radius of a non-rotating star of the same mass and metallicity (see Appendix~\ref{sec:che_thresholds}).   We compute the mass loss rate for \ac{CH} stars in the same way as for regular \ac{MS} stars, but with this fixed rather than evolving radius.  As a consequence, the total mass lost over the \ac{MS} by \ac{CH} stars in our \ac{COMPAS} models is generally within $\lesssim 10\%$ of that lost by non-\ac{CH} stars of the same \ac{ZAMS} mass and metallicity, except for the most massive stars in our simulations, with initial masses above $100\, M_\odot$, where the absence of radial expansion leads to significantly reduced \ac{MS} mass loss estimates for \ac{CH} stars.  Finally, we assume that if a star evolves chemically homogeneously through the main sequence, it contracts directly into a naked helium star at the end of the main sequence, retaining its full mass at that point.  Future evolution follows the \citet{Hurley_2000} models of helium stars.  

Tides are very efficient at ensuring circularisation and synchronisation in very close binaries through tidal locking \citep[e.g.,][]{Hut:1981}.  We therefore assume that all potential candidates for \ac{CHE} are tidally synchronised at birth, so that their rotational angular frequency equals the orbital angular frequency.   We check this angular frequency at birth to determine whether a star belongs to the \ac{CH} type and continue to check it at every time step on the main sequence.  If the angular frequency ever drops below the threshold value for \ac{CHE}, e.g., because of binary widening as a consequence of mass loss through winds, the star is henceforth evolved as a regular main sequence star (in our simplified treatment, it immediately jumps to the track of a regular main sequence star of the same mass).  We assume that once a chemical gradient is formed, it is very challenging to overcome and ensure efficient mixing, so in our model, a star that is not evolving chemically homogeneously cannot become a \ac{CH} star (cf.~BPASS models, which allow quasi-chemically homogeneous evolution through accretion-induced spin-up \citealt{Eldridge:2017}).  Although we assume perfect tidal synchronisation for \ac{CH} stars, we disregard the angular momentum stored in the stellar rotation when considering binary evolution with mass loss.

\subsubsection{Initial conditions}\label{sec:COMPASphysics_initial}

Each binary system in a \ac{COMPAS} simulation is described at birth (i.e. at \acf{ZAMS}) by its initial conditions: constituent star masses, separation, eccentricity and metallicity.  Initial conditions for our experiments were chosen using statistical distribution functions from the literature that were themselves based on observations.  We describe the most important of these, and some important parameters that affect the evolution of the constituent stars as well as the binary system, in the following paragraphs. 

The mass of the primary star in the binary system (the more massive star at \ac{ZAMS}) \monei is described by the \citet{Kroupa_2000} \ac{IMF}, the distribution function of which is given by 
\begin{equation}
    p(m_{1,i})\propto m_{1,i}^{-\alpha},
\end{equation}
where $\alpha=2.3$ for the simulated range of primary masses $\monei\in[5,150]$\Msun.  We assume that the \ac{IMF} is the same for all metallicities. 

The mass of the secondary star (less massive at \ac{ZAMS}) \mtwoi is determined by drawing a mass ratio between the constituent stars $\qi~\equiv~\mtwoi~/~\monei$ that follows a flat distribution, $p(q_i)=1$ \citep{Sana_2012, Kobulnicky_2014}.  Since we are interested in \ac{BBH} formation, we  explore only \mtwoi\nolinebreak$\geq$\nolinebreak3.0\Msun here.   However, for both the primary and secondary mass, we consider the full mass range to normalise the simulation results to a given star-forming mass or star formation rate \citep[e.g.,][]{Neijssel_2019}. 

The initial separation is drawn from a flat-in-log distribution independently of the masses (see \citealt{MoeDiStefano:2017} for coupled initial conditions):
\begin{equation}
    p(a_i)\propto\frac{1}{a_i},
\end{equation}
where $a_i~\in~[0.01, 1000]$\AU~\citep{Opik_1924, Abt_1983}.

We assume all binaries are circular at birth (i.e.~$\ei\nolinebreak=\nolinebreak0$), see \citet{VignaGomez:2020} for further discussion.  Close binaries are tidally circularised at birth, so this has no impact on potential \ac{CHE} systems.  

We simulate thirty different metallicities spaced uniformly in the logarithm across the range 
$\minus4\leq\logten{Z}\leq\minus1.825$.

\subsubsection{Wind-driven mass loss}\label{subsubsec:COMPASphysics_mass_loss}

We use the mass loss rates as prescribed by \citet[]{Hurley_2000, Hurley_2002} and references therein for cooler stars with temperatures of $12,500$\Kelvin and below. For stars hotter than $12,500$\Kelvin we use the wind mass loss rates from \citet{Vink_2001}, as implemented in \citet{Belczynski_2010}.

The \ac{LBV} stars \citep{Maeder_1989, Pasquali_1997}, located close to the Humphreys-Davidson limit in the \ac{HRD} \citep{Humphreys_1979}, are treated differently. For these stars we use the \ac{LBV} wind mass loss rate prescribed by \citet{Belczynski_2010}:

\begin{equation}
    \dMdt=f_\mathrm{LBV}\tenpow{\minus4}\Msun yr^{\minus1},
\end{equation}{\textnormal{where}\ $f_\mathrm{LBV}=1.5$}.

For massive, hot and bright naked helium stars, we use a metallicity-dependent \ac{WR} wind mass loss rate \citep{Vink_2005}. We parametrise the rate of mass loss by following \citet{Belczynski_2010}:

\begin{equation}
    \dMdt=f_\mathrm{WR}\tenpow{\minus13}{L^{1.5}}{(\frac{Z}{\Zsun})^m}\Msun yr^{\minus1},
    \label{eqn:WRwinds}
\end{equation}
where $L$ is the luminosity, $m=0.86$ \citep{Vink_2005}, we take $Z_\odot=0.014$ \citep{Asplund:2009} and $f_\mathrm{WR}=1.0$ in our default model. 

In our model all stars that remain chemically homogeneous on the main sequence convert their entire mass into helium at the end of their main sequence lifetime, so all such stars evolve into naked helium stars. Lower mass loss rates would promote the formation of black holes as the end products of the evolution of these massive stars, so we consider four different values of $f_\mathrm{WR}$ in order to study the impact of \ac{WR} mass loss on \ac{CHE}: $ f_\mathrm{WR}\in\{\,0.0\,, 0.2\,, 0.6\,, 1.0\,\}$.

All mass lost in winds is assumed to promptly depart the binary without further interaction with the companion in so-called `Jeans mode' mass loss, carrying away the specific angular momentum of the donor.  

Our mass loss rate models do not include the impact of stellar rotation, which is likely to drive additional mass loss.  In particular, the \ac{WR} star formed when the \ac{CH} star contracts and spins up at the end of its \ac{MS} is likely to be critically rotating, and the same process may repeat during core contraction after core helium depletion.  The star's angular momentum can be reduced to sub-critical levels with a small amount of mass loss, and therefore does not significantly impact the overall mass budget; however, this does affect the remnant spin \cite{MarchantMoriya:2020}.

\subsubsection{Mass transfer and over-contact systems}
\label{sec:COMPASphysics_mass_transfer}

We use the prescriptions described in \citet{COMPAS_VignaGomez_2018,Neijssel_2019} to determine  the dynamical stability of mass transfer through \ac{RLOF}, the fraction of mass accreted onto the companion and the specific angular momentum carried away by non-conservative dynamically stable mass transfer, and the outcome of common-envelope evolution.  For non-\ac{CHE} binaries that go through a common-envelope phase, we assume that Hertzsprung-gap donors do not survive (the 'pessimistic' prescription of \citealt{Belczynski:2007,Neijssel_2019}) and we assume that immediate post-common-envelope \ac{RLOF} indicates a merger.

We deviate from previous \ac{COMPAS} models in the treatment of binaries that experience \ac{RLOF} at \ac{ZAMS}.  Unlike previous work, we now allow such binaries to equilibrate their masses.  The new separation of the equal-mass binary with a conserved total mass is determined by angular momentum conservation.  Binary components are allowed to over-fill their Roche lobes, creating over-contact systems.  However, if the components extend past the L2 Lagrange points after equilibration, we assume that the binary loses co-rotation and promptly merges \citep{Marchant_2016}.  For equal-mass circular binaries, the volume-equivalent radius for half of the volume within the L2 equipotential surface equals half the orbital separation. Therefore, our criterion for avoiding a prompt merger is equivalent to demanding that the sum of the unperturbed stellar radii is smaller than the orbital separation $a$.

\subsubsection{Pair-instability supernovae}\label{sec:COMPASphysics_pisn}

Stellar evolution models predict that single stars with helium cores in the range $\sim 60$\,--\,130\,\Msun can become unstable due to electron-positron pair production, leading to \acp{PISN} which disrupt the star, leaving no remnant behind \citep[e.g.,][]{Fowler_1964, Barkat_1967, Fraley_1968, Woosley_2002,Woosley_2019, Farmer_2019}.   Stars with helium cores more massive than $130\, M_\odot$ also experience a rapid collapse driven by pair production, but in these stars photodisintegration prevents a subsequent explosion; such stars may again produce merging \acp{BBH} \citep[e.g.,][]{Marchant_2016,duBuisson_2020}, but are not explored in our models, which have maximum initial stellar masses of $150\,M_\odot$.  Meanwhile, stars with somewhat lower helium core masses, between $\sim 35$ and $\sim 60$ \Msun, are predicted to eject significant fractions of their total mass over several episodes \citep[e.g.,][]{Yoshida_2016,Woosley_2017,Marchant_2019, Renzo_2020}.  Such \acp{PPISN} leave behind a black hole remnant, albeit with a reduced mass.  \acp{PISN} and \acp{PPISN} are expected to produce a \ac{PISN} mass gap in the distribution of remnant masses from single stellar evolution -- a dearth of black holes with masses between $\sim 45 M_\odot$ and $\sim 130 M_\odot$.

Some superluminous supernovae have been identified as \ac{PISN} candidates \citep[][and references therein]{GalYam:2012}, while iPTF2014hls has been identified as a \ac{PPISN} candidate \citep{Arcavi:2017}.  Furthermore, the distribution of masses of gravitational-wave observations appeared consistent with a cutoff due to (P)PISNe \citep{Abbott_2019}, though GW190521 is a \ac{BBH} merger with at least one component in the predicted \ac{PISN} mass gap \citep{GW190521}.   

Here, we follow the \citet{Stevenson:2019} fit to the \citet{Marchant_2019} models for predicting the range of \ac{PISN} masses and the \ac{PPISN} remnant masses from the masses of the progenitor helium cores.  We apply the entire \ac{PPISN} mass loss in one time step.  Moreover, in our treatment both supernovae happen in one timestep for equal-mass stars.  This over-estimates the post-supernova period and  eccentricity of binaries whose components lose significant mass in a \ac{PPISN}.

We use the 'Delayed' prescription of \citet{Fryer:2012} for compact object remnant masses and modulate the natal kicks by fallback for regular core-collapse supernovae, with reduced kicks for electron-capture and ultra-stripped supernovae as in \citet{COMPAS_VignaGomez_2018}.

\subsection{Star formation rate}\label{subsec:SFR}

The local merger rate of \acp{BBH} depends on their formation rate at higher redshifts due to the possibly significant time delays between formation and merger, and is therefore sensitive to the star formation rate as a function of redshift.  Furthermore, the yield of \acp{BBH} per unit star forming mass, the \ac{BBH} mass distribution, and the distribution of delay times between formation and merger are all sensitive functions of the metallicity of progenitor stars, both for \ac{CHE} \citep[e.g.][]{Marchant_2016} and non-\ac{CHE} \citep{Neijssel_2019, Chruslinska:2019} systems.  We must therefore specify a \ac{MSSFR} in order to estimate the merger rate and properties of \acp{BBH}.  We use the preferred model of \citet{Neijssel_2019} for the \ac{MSSFR}.  Figure~\ref{fig:madau_plot} shows the contribution of different ranges of star formation metallicities to the total star formation rate.  This model has higher star formation metallicities in the local Universe than the \citet{Taylor_2015} model used by \citet{duBuisson_2020} (cf.~their Figure 2).

\bigskip\noindent
\begin{minipage}{\linewidth}
    \includegraphics[viewport = 5 0 647 505, width=8.45cm, clip]{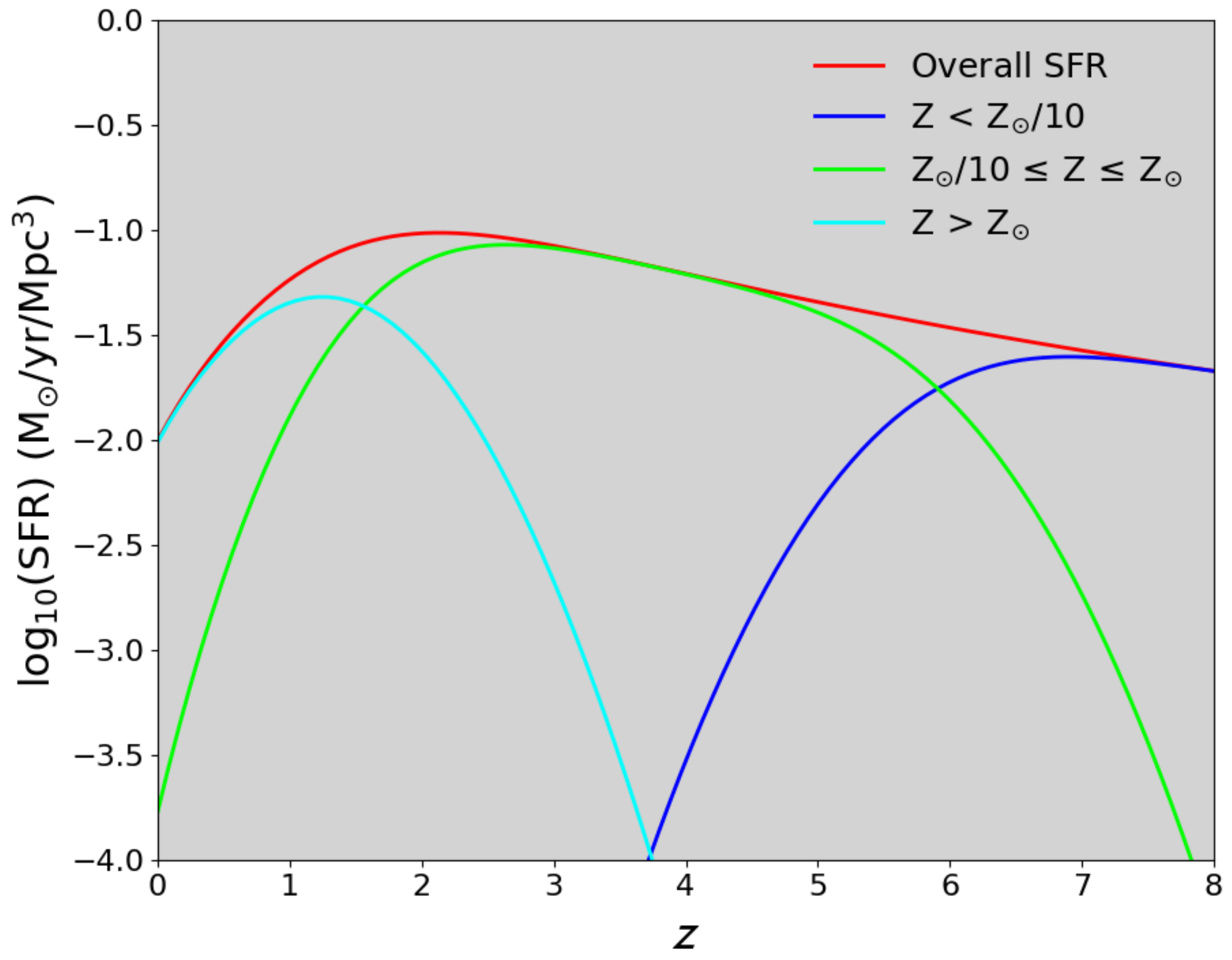}
    \captionof{figure}{The total star formation rate as a function of redshift (red) and subdivided into different ranges of metallicity, following the preferred model of \citet{Neijssel_2019}. The dark blue and green curves are most relevant for \ac{BBH} formation.}
    \label{fig:madau_plot}
\end{minipage}

\section{Results and discussion}\label{sec:results}

We evolved a total of $12$ million binaries as described in section~\ref{sec:methods}.  These were equally divided into 30 metallicity bins and 4 choices of the \ac{WR} mass loss rate multipliers $f_\mathrm{WR}$, for a total of $100,000$ binaries for each of 120 combinations of $Z$ and $f_\mathrm{WR}$.  Binaries are evolved until a double compact object is formed, or until an event happens which makes this outcome impossible (e.g., the stars merge or the binary becomes unbound),  or the system reaches 14 Gyr in age.

Our simulations are based on a Monte Carlo sampling of binaries. We estimate the sampling uncertainty on all derived quantities via bootstrapping: we uniformly resample, with replacement, a new population of $12$ million binaries from the original, evolved, population of $12$ million. Error bars on plots, where shown, correspond to the 5\tsup{th} and 95\tsup{th} percentiles from bootstrapping.

\subsection{Population statistics}\label{subsec:results_population_stats}

\begin{table*}
    \centering
    \caption{Population Statistics}
    \begin{tabular}{|E{6.65cm}|G{1.63cm}|G{1.63cm}|G{1.63cm}|G{1.63cm}|G{1.8cm}|}
        \hline
        \multicolumn{6}{|c|}{\cellcolor{lightgray}Population} \\
        \hline
        & $f_\mathrm{WR}=0$ & $f_\mathrm{WR}=0.2$ & $f_\mathrm{WR}=0.6$ & $f_\mathrm{WR}=1.0$ & \textit{Total} \\
        \hline
        Number of binaries evolved & 3,000,000 & 3,000,000 & 3,000,000 & 3,000,000 & 12,000,000 \\
        \hline
        L2 overflow at ZAMS & 243,717 & 243,638 & 243,419 & 243,930 & 974,704 \\
        \hline
        Surviving binaries & 2,756,283 & 2,756,362 & 2,756,581 & 2,756,070 & 11,025,296 \\
        \hline
        \multicolumn{6}{|c|}{\cellcolor{lightgray}Surviving Population} \\
        \hline
        At least one star experiencing RLOF at ZAMS & 75,530 & 75,410 & 75,328 & 75,707 & 301,975 \\
        \hline
        Both stars in binary \acs{CH} at \acs{ZAMS} & 4,193 & 4,281 & 4,201 & 4,216 & 16,891 \\
        \hline
        Primary only \acs{CH} at \acs{ZAMS} & 2,607 & 2,615 & 2,593 & 2,604 & 10,419 \\ 
        \hline
        Secondary only \acs{CH} at \acs{ZAMS} & 0 & 0 & 0 & 0 & 0 \\
        \hline
        Post-ZAMS Merger & 618,001 & 616,749 & 618,425 & 620,250 & 2,473,425 \\
        \hline
        BBHs formed & 68,231 & 67,200 & 66,016 & 60,294 & 261,741 \\
        \hline
        BBHs merging in 14 Gyr & 11,004 & 11,048 & 10,926 & 10,647 & 43,625 \\
        \hline
        \multicolumn{6}{|c|}{\cellcolor{lightgray}Both stars \acs{CH} at \acs{ZAMS}} \\
        \hline
        At least one star experiencing RLOF at ZAMS & 3,661 & 3,761 & 3,715 & 3,715 & 14,852 \\
        \hline
        Both stars remained \acs{CH} on \acs{MS} & 3,444 & 3,461 & 3,379 & 3,360 & 13,644 \\ 
        \hline
        Primary only remained \acs{CH} on \acs{MS} & 43 & 89 & 116 & 160 & 408 \\ 
        \hline
        Secondary only remained \acs{CH} on \acs{MS} & 0 & 0 & 0 & 0 & 0 \\ 
        \hline
        Neither star remained \acs{CH} on \acs{MS} & 706 & 731 & 706 & 696 & 2,839 \\ 
        \hline
        BBHs formed & 2,152 & 2,370 & 2,527 & 2,621 & 9,670 \\ 
        \hline
        BBHs merging in 14 Gyr & 2,057 & 2,322 & 2,377 & 2,306 & 9,062 \\ 
        \hline
        \multicolumn{6}{|c|}{\cellcolor{lightgray}Primary only \acs{CH} at \acs{ZAMS}} \\
        \hline
        At least one star experiencing RLOF at ZAMS & 0 & 0 & 0 & 0 & 0 \\
        \hline
        Primary remained \acs{CH} on \acs{MS} & 1,405 & 1,353 & 1,341 & 1,337 & 5,436 \\ 
        \hline
        BBHs formed & 0 & 2 & 3 & 7 & 12 \\ 
        \hline
        BBHs merging in 14 Gyr & 0 & 0 & 0 & 0 & 0 \\ 
        \hline
    \end{tabular}
    \label{table:pop_stats}
\end{table*}

The population statistics are shown in Table~\ref{table:pop_stats}. From a population of 11,025,296 binaries that survived beyond \ac{ZAMS} (i.e. did not merge at \ac{ZAMS}), 16,891 were composed of two \ac{CH} stars at \ac{ZAMS}, with a further 10,419 composed of one \ac{CH} star and one \ac{MS} star at \ac{ZAMS}. Furthermore, in all of the binaries with only one \ac{CH} star at \ac{ZAMS} it was, as we would expect, the primary, more massive, star that was \acl{CH}.  A total of 261,741 \acp{BBH} were formed in the simulation, but only 43,625 of these were close enough to merge within 14 Gyr, the current age of the Universe.  Among the 13,644 simulated binaries that evolved chemically homogeneously throughout the main sequence, 9,670 went on to form \acp{BBH}, the vast majority of which, 9,062, merged within 14 Gyr (the few non-merging ones are those which lost significant mass in \acp{PPISN}). 

\subsection{Evolved system properties}\label{subsec:results_evolved_properties}

\begin{figure*}
    \begin{center}
        \hspace{-1.0cm}
	    \includegraphics[viewport = 0 0 125 75, width = 15.75cm, clip]{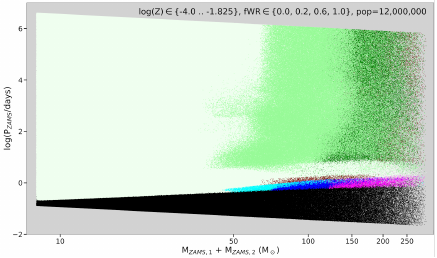}
	    \vspace{-1.0mm}
        \includegraphics[viewport = 0 0 1675 110, width=12.0cm, clip]{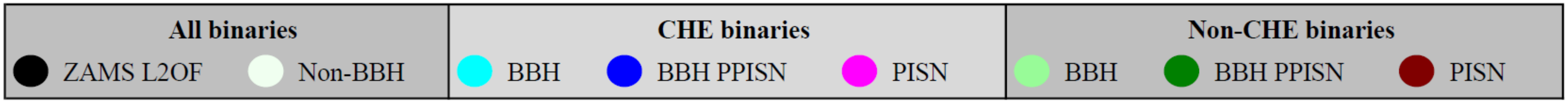}
    \end{center}
    \vspace{-2.00mm}
    \captionof{figure}{Initial parameters and final outcomes for each of the binary systems synthesised, showing the initial orbital period $T_\mathrm{ZAMS}$ (in days) vs the initial total mass (in \Msun). The population represents a grid of 30 metallicities evenly spaced over the range \mbox{-4~$\leq\,\log_{10}(Z)\,\leq$~-1.825}, and 4 \ac{WR} mass loss multipliers, $f_\mathrm{WR}\in{\{}\,0.0\,, 0.2\,, 0.6\,, 1.0\,{\}}$. Regions shaded in black represent all systems that experienced L2 overflow at ZAMS; pale green represents systems that did not form \acs{BBH}s. Systems for which both stars were chemically homogeneous at \ac{ZAMS} and remained so throughout their main sequence lifetime are represented by regions shaded cyan if they formed \ac{BBH}s via regular core-collapse supernovae, blue if they formed \ac{BBH}s after undergoing \acp{PPISN}, and magenta if they exploded as \acp{PISN}. Systems in which at least one of the stars did not evolve chemically homogeneously for its entire main sequence lifetime are represented by areas shaded light green if they formed \acp{BBH} via core-collapse supernovae, dark green if they formed \acp{BBH} following \acp{PPISN}, and maroon if either star exploded as a \ac{PISN}.
    }
    \label{fig:chescatterplot_all}
\end{figure*}

Figure~\ref{fig:chescatterplot_all} presents a visual summary of the evolutionary outcomes for each of the 12 million binary systems synthesised, with each point on the plot representing a single binary system, and the colour indicating the initial parameters and the outcome of the evolution (per the legend). We are particularly interested in systems for which both stars evolve chemically homogeneously and eventually collapse to form a \ac{BBH}, so we have agglomerated some of the less interesting progenitor types and outcomes into groups so that the plot is not overly busy. Because Figure~\ref{fig:chescatterplot_all} is a summary over the entire grid of metallicities and \ac{WR} mass loss rate multipliers synthesised, it allows us to see on a broad scale the evolutionary outcomes for both \ac{CHE} systems and non-\ac{CHE} systems. The COMPAS models for the formation of non-\ac{CHE} \acp{BBH} have been discussed by \citep{Neijssel_2019}, so we will focus our discussions hereafter one the \ac{CHE} channel.

\begin{figure*}
    \noindent\centering
    \includegraphics[viewport = 5 0 850 560, width=12.5cm, clip]
    {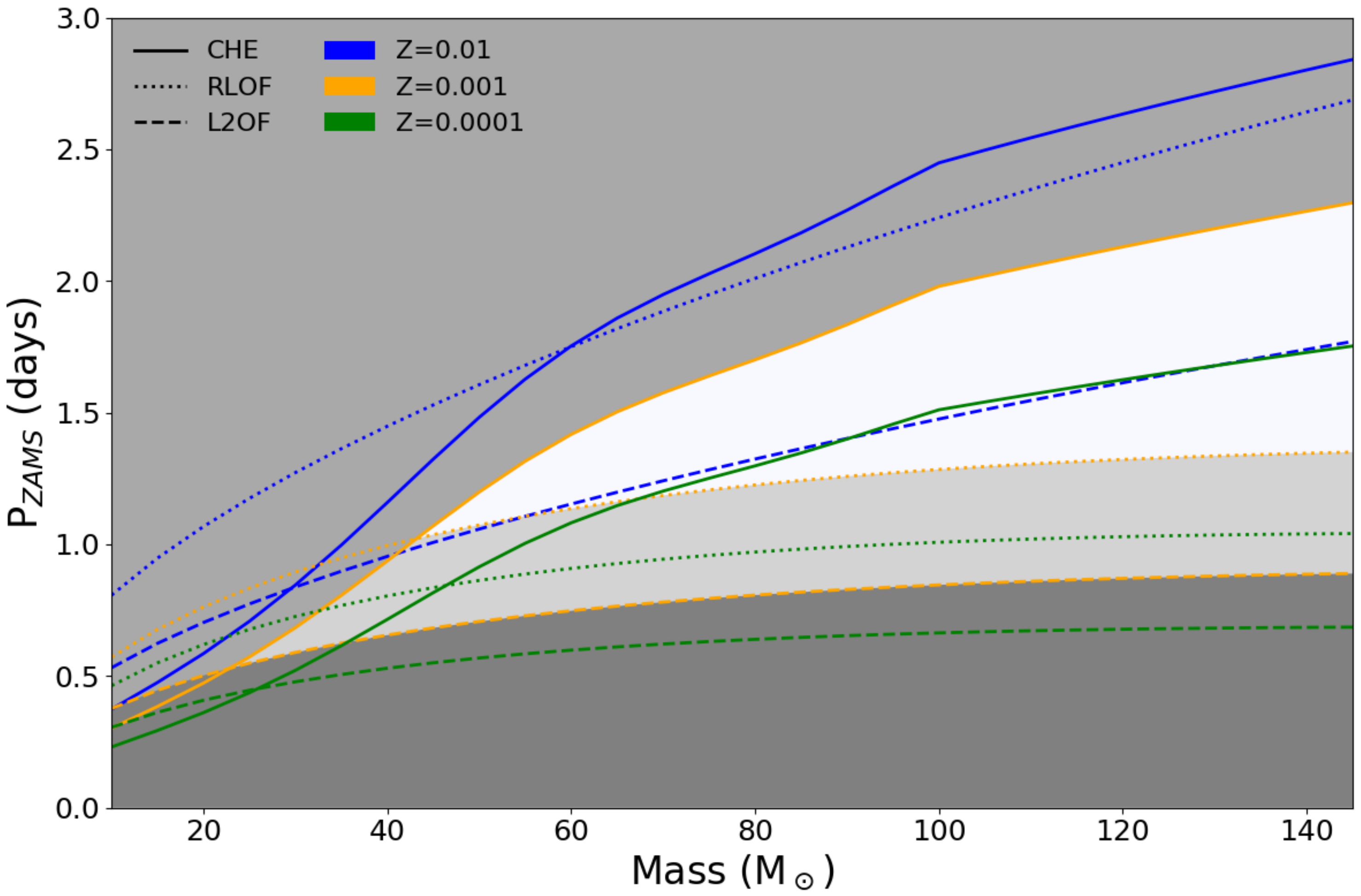}
    \captionof{figure}{Parameter space for equal-mass binary systems with the indicated companion mass at which chemically homogeneous evolution is expected to occur at \ac{ZAMS}.  Solid lines show the thresholds for \ac{CHE} implemented in COMPAS (see Appendix~\ref{sec:che_thresholds}), dotted lines are \ac{RLOF} thresholds, and dashed lines are L2 overflow thresholds. Colours differentiate metallicities.  Shading corresponds to $Z=0.001$: the dark colour at the bottom indicates L2 overflow at \ac{ZAMS}, grey at the top indicates periods too low for \ac{CHE} at \ac{ZAMS}, while the region below the solid line and above the dashed lined indicates the possible range for \ac{CHE}, with the parameter space for over-contact binaries that undergo \ac{RLOF} at \ac{ZAMS} shaded more darkly.  
    }
    \label{fig:CHEBoundaries}
\end{figure*} 

Figure~\ref{fig:CHEBoundaries} shows the parameter space in which \ac{CHE} is expected to occur in synchronously rotating binaries according to our \ac{CHE} threshold. The darkest grey area in the lower part of the diagram indicates the region in which L2 overflow occurs and the stellar components merge; the lighter grey area in the upper part indicates the region in which the stellar components do not rotate rapidly enough to induce \ac{CHE}. The central, lighter, area of the diagram indicates the region in which we expect \ac{CHE} to occur, with the darker, lower, part of the central area indicating the important region of binaries whose components overflow their Roche lobes but avoid L2 overflow, occupied by the over-contact systems described by \citet{Marchant_2016}.  This over-contact region is responsible for much of the \ac{BBH} formation through \ac{CHE} (cf.~Figure~\ref{fig:chescatterplot_all}).

As expected (given our \ac{PPISN} and \ac{PISN} mass limits, see section~\ref{sec:COMPASphysics_pisn}), we see \acp{BBH} from \acp{PPISN} begin to appear at a total \ac{ZAMS} mass of $\gtrsim70$\,\Msun while \ac{PISN} events appear at a total \ac{ZAMS} mass of $\gtrsim120$\,\Msun.  A few unbound \ac{CHE} systems correspond to simultaneous \acp{PISN} that instantaneously removed more than half the mass of the binary in our treatment (see Section \ref{sec:COMPASphysics_pisn}).  In practice, such systems will undergo a series of pulsations leading to non-simultaneous mass loss and may survive, but at separations too large to merge within the current age of the Universe.  The horizontal band of \acp{PISN} just above the \ac{CH} binaries in Figure~\ref{fig:chescatterplot_all} are hybrid systems comprised of a \ac{CH} star and a \ac{MS} star, whereas the vertical band of \acp{PISN} at the upper right of the plot are systems comprised of two \ac{MS} stars.

\subsection{Population synthesis}\label{subsec:pop_synth}

\begin{figure}
    \centering
    \includegraphics[viewport = 0 5 680 615, width=8.45cm, clip]{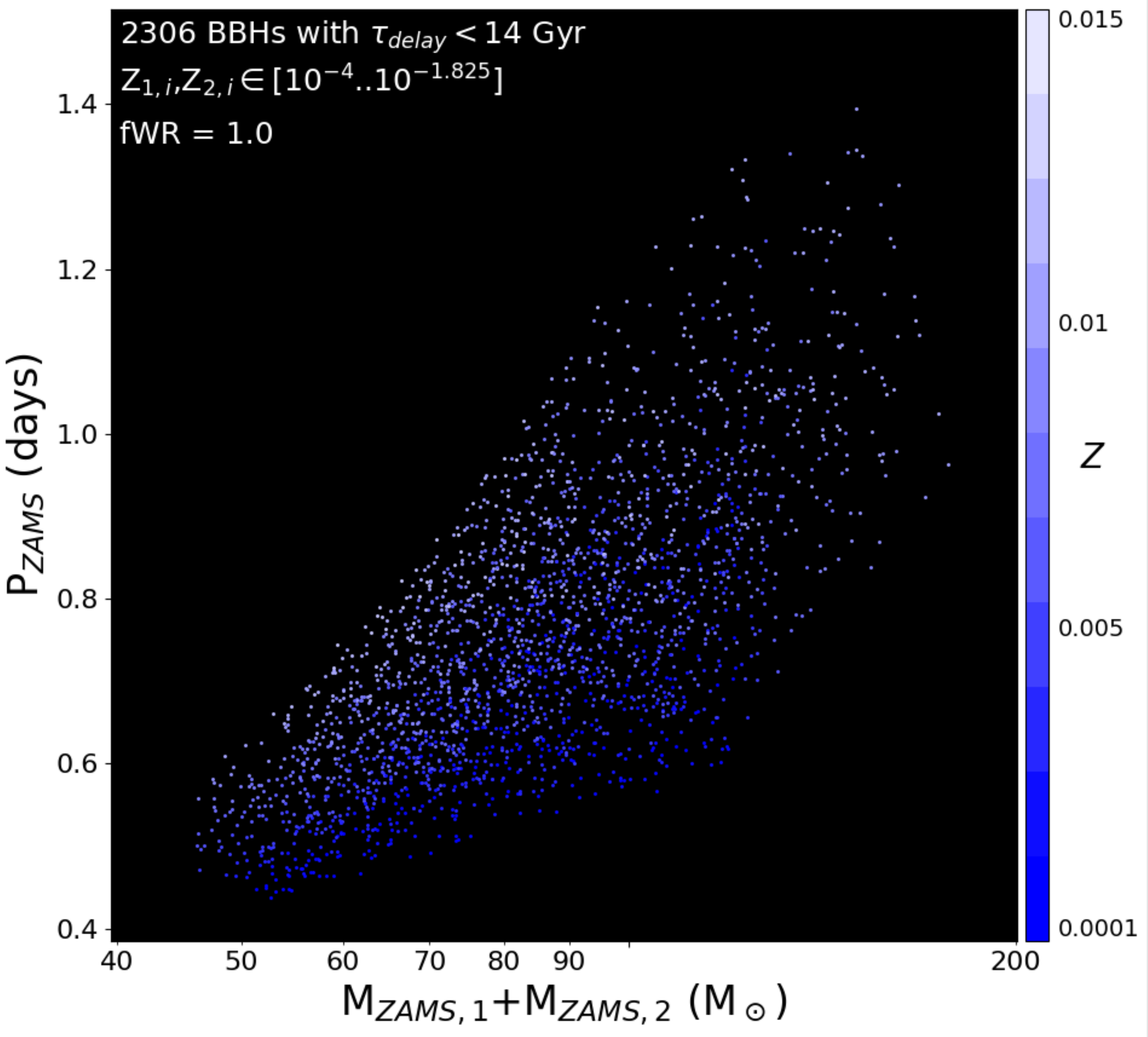}
    \caption{Initial total masses and orbital periods for \ac{CHE} systems that go on to form \acp{BBH} that will merge within 14 Gyr. Each point represents one simulated binary, evolved with \ac{WR} mass loss multiplier $f_\mathrm{WR}=1$, shaded according to its metallicity.
    }
    \label{fig:pInitialVSmTotal}
\end{figure}

The initial system total masses and orbital periods of \ac{CHE} systems that go on to form \acp{BBH} merging within 14 Gyr are shown in Figure~\ref{fig:pInitialVSmTotal}.  We show binaries evolved with the \ac{WR} mass loss multiplier $f_\mathrm{WR}=1$.  Each point on the plot represents a simulated binary shaded according to its metallicity.  Higher metallicity binaries are shifted toward the top of the plot.  This is consistent with Figure \ref{fig:CHEBoundaries}, which shows that higher-metallicity stars have greater stellar radii and hence greater minimal separation, as well as lower \ac{CHE} threshold rotational frequency.  

Binaries with reduced \ac{WR} winds have similar initial distributions, but show a clear-cut maximum total mass of $\approx 120\,M_\odot$, which matches the mass threshold of $60\,M_\odot$ for individual He star masses beyond which \acp{PISN} occur and leave no remnants.  At higher $f_\mathrm{WR}$, high-metallicity systems can lose a significant fraction of their mass, so binaries with initial total masses above $120\,M_\odot$ can avoid \acp{PISN}.  

\begin{figure}
  \centering
    \includegraphics[width=8.45cm, clip]{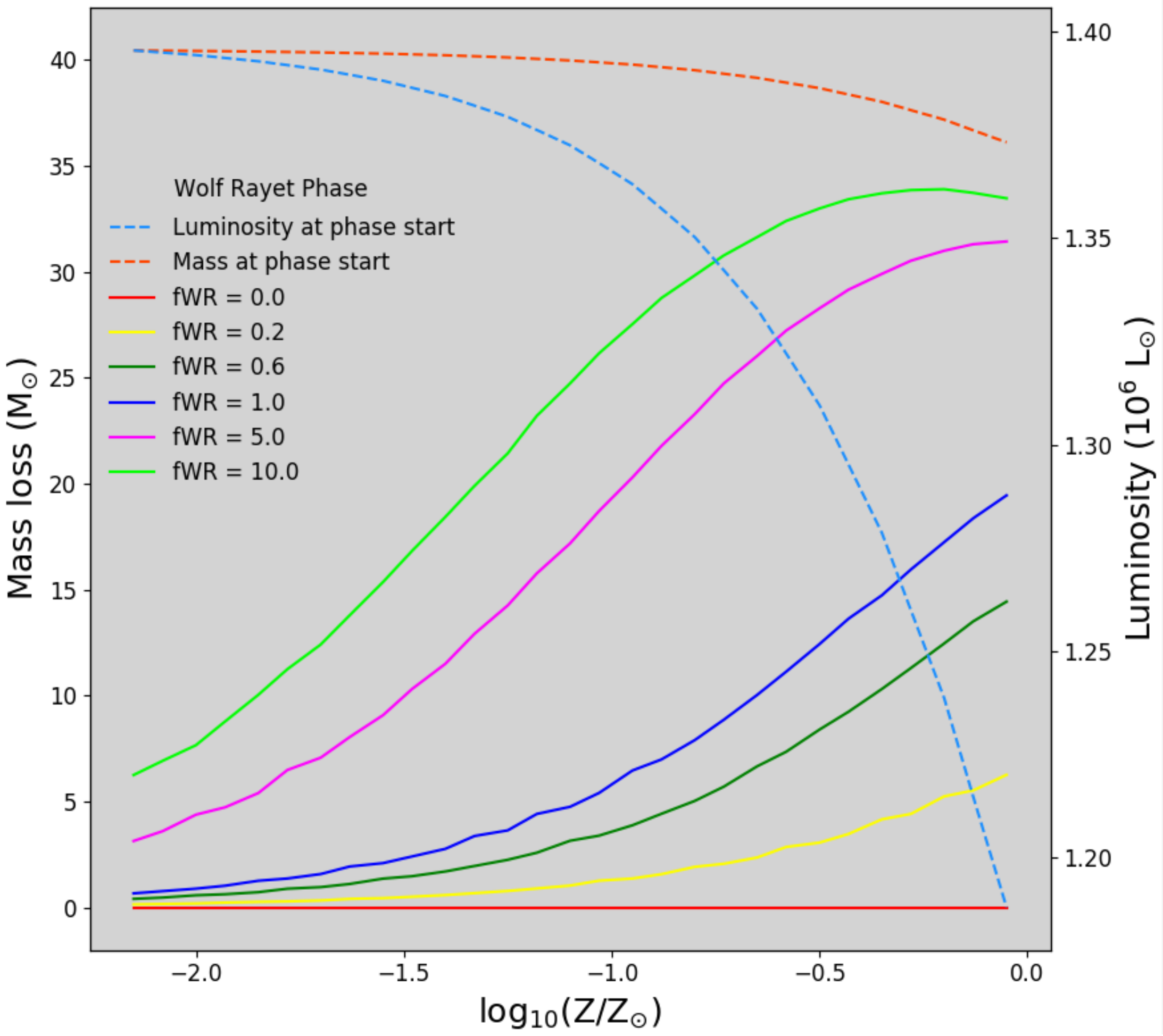}
    \captionof{figure}{Total mass lost by a \ac{WR} star with a \ac{ZAMS} mass of $40.5 \Msun$ as a function of metallicity. Line colour indicates the \ac{WR} mass loss rate multiplier (solid lines). Also shown are the mass (on the same scale as the mass loss curves) and luminosity at the start of the \ac{WR} phase as a function of metallicity (dashed lines).}
    \label{fig:WRActualMassLoss}
\end{figure}

To illustrate this, we plot the mass lost by a \ac{CH} star with a \ac{ZAMS} mass of $40.5\,M_\odot$ over the naked helium phase in Figure \ref{fig:WRActualMassLoss}, for a range of \ac{WR} mass loss multipliers and metallicities.  At $f_\mathrm{WR}=1$ and $Z=Z_\odot$, this star loses nearly half of its mass in \ac{WR} winds.  Meanwhile, at low metallicities, which are typical for high formation redshifts, the total mass lost in \ac{WR} winds is very low, except at artificially enhanced $f_\mathrm{WR}$ values of 5 and 10, which disagree with observational constraints and are not considered in this study.  Consequently, we do not expect to see a significant impact of $f_\mathrm{WR}$ on low-metallicity \ac{BBH} formation, which matches our findings as discussed below.

Table~\ref{table:pop_stats} shows, across all simulated metallicities and \ac{WR} mass loss multipliers,  $\sim80\%$ of binaries composed of two \ac{CH} stars at \ac{ZAMS} retain two \ac{CH} stars at the end of the main sequence.  For binaries composed of one \ac{CH} star and one \ac{MS} star at \ac{ZAMS}, the \ac{CH} star will remain chemically homogeneous by the end of the \ac{MS} in only $\sim 50\%$ of simulations.   Since we assume tidal locking in the \ac{CHE} model implemented in \ac{COMPAS}, as a binary widens due to mass loss and the orbital frequency of the binary slows, the rotational frequency of the constituent \ac{CH} stars slows commensurably.  Binaries in which only the primary is \ac{CH} at \ac{ZAMS} avoided \ac{RLOF} and are typically wider, so further widening through winds is more likely to spin down the primary sufficiently to evolve off the \ac{CHE} track.

\begin{figure}
    \centering
    \includegraphics[width=8.45cm, clip]{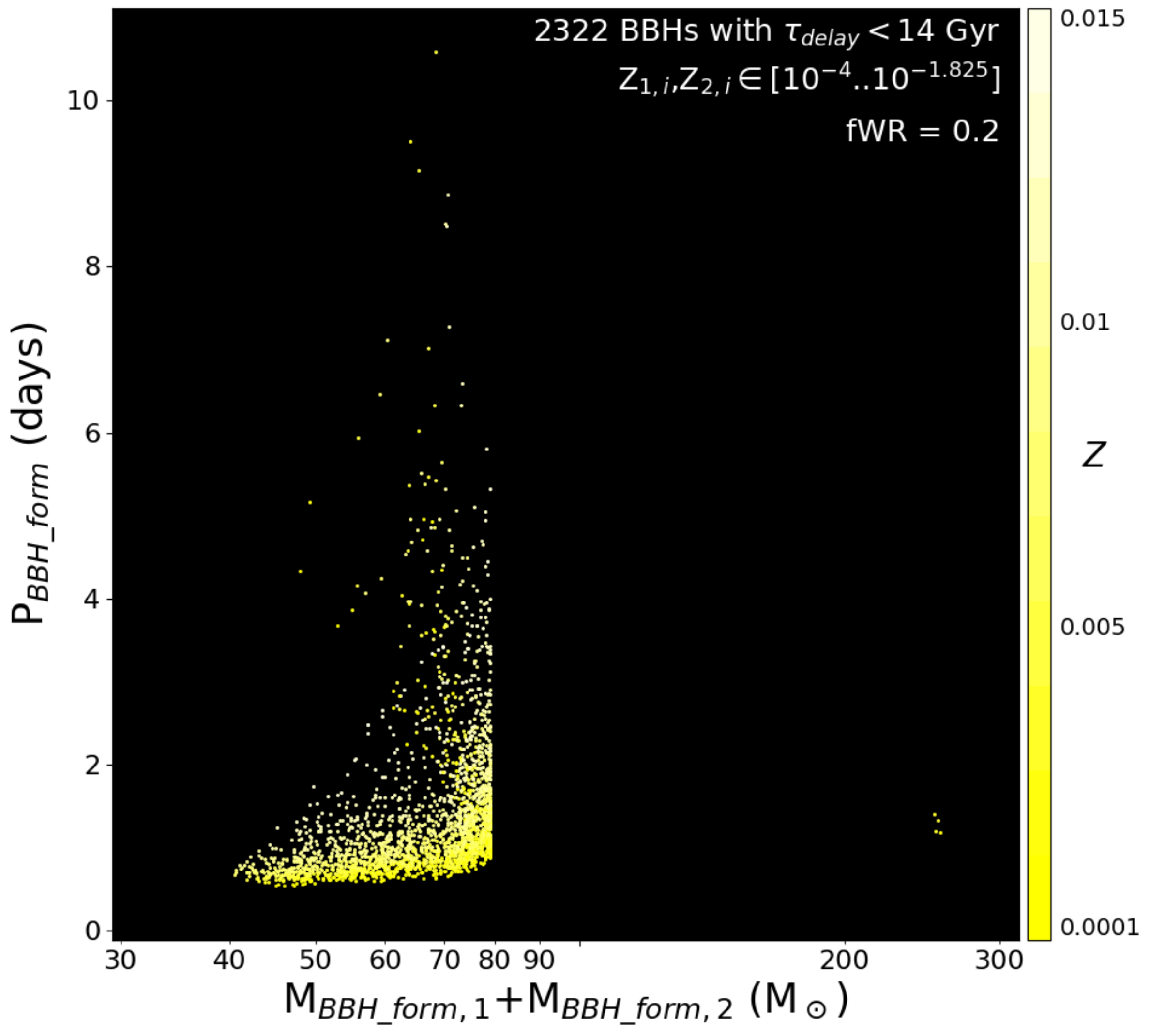}
    \caption{Total masses and orbital periods immediately after \ac{BBH} formation for \ac{CHE} systems that will merge within 14 Gyr.  Each point represents a simulated binary, evolved with \ac{WR} mass loss multiplier $f_\mathrm{WR}=0.2$, shaded according to its metallicity. The empty area between $\approx 80$\,\Msun and $\approx 250$\,\Msun is a consequence of systems that lost mass as \acp{PPISN} or left no remnants after exploding as \acp{PISN}.}
    \label{fig:pFinalVSmTotal}
\end{figure}

Figure~\ref{fig:pFinalVSmTotal} shows the distribution of the \ac{BBH} total masses and orbital periods just after \ac{BBH} formation for systems evolving through the \ac{CHE} channel.  As in Figure~\ref{fig:pInitialVSmTotal}, we select only \acp{BBH} that will merge in 14 Gyr and shade binaries by metallicity.  On this plot, we select $f_\mathrm{WR}=0.2$.  This allows us to show not only the sharp disappearance of \acp{BBH} with total masses above $\approx 80\, M_\odot$ due to \ac{PPISN} mass loss and complete disruption in \acp{PISN}, but also their reappearance at masses above $\approx 250\, M_\odot$, on the other side of the `\ac{PISN} mass gap'. There are only very few such high-mass binaries in our simulations because, with our \ac{ZAMS} mass upper limit of $150\,M_\odot$, they require very low mass loss.  Consequently, there are no such binaries in our $f_\mathrm{WR}=1.0$ simulations because their progenitors lose too much mass to remain above the \ac{PISN} threshold.

The shortest post-\ac{BBH} formation periods, and thus the shortest delay times, are seen for the lowest-metallicity systems.  This is due to the combined effects of their lower period at \ac{ZAMS} as seen in Figure \ref{fig:pInitialVSmTotal} and the reduced orbital widening due to reduced mass loss at low metallicities.  However, some low-metallicity binaries lose sufficient mass in \acp{PPISN} to create wider, more eccentric binaries found toward the top of Figure~\ref{fig:pFinalVSmTotal}.

\subsection{Binary Black Holes}\label{subsec:results_BH_mergers}

\subsubsection{Formation rates}
\label{subsubsec:results_formation_rates}

\begin{figure}
    \centering
    \includegraphics[viewport = 0 0 550 490, width=8.45cm, clip]{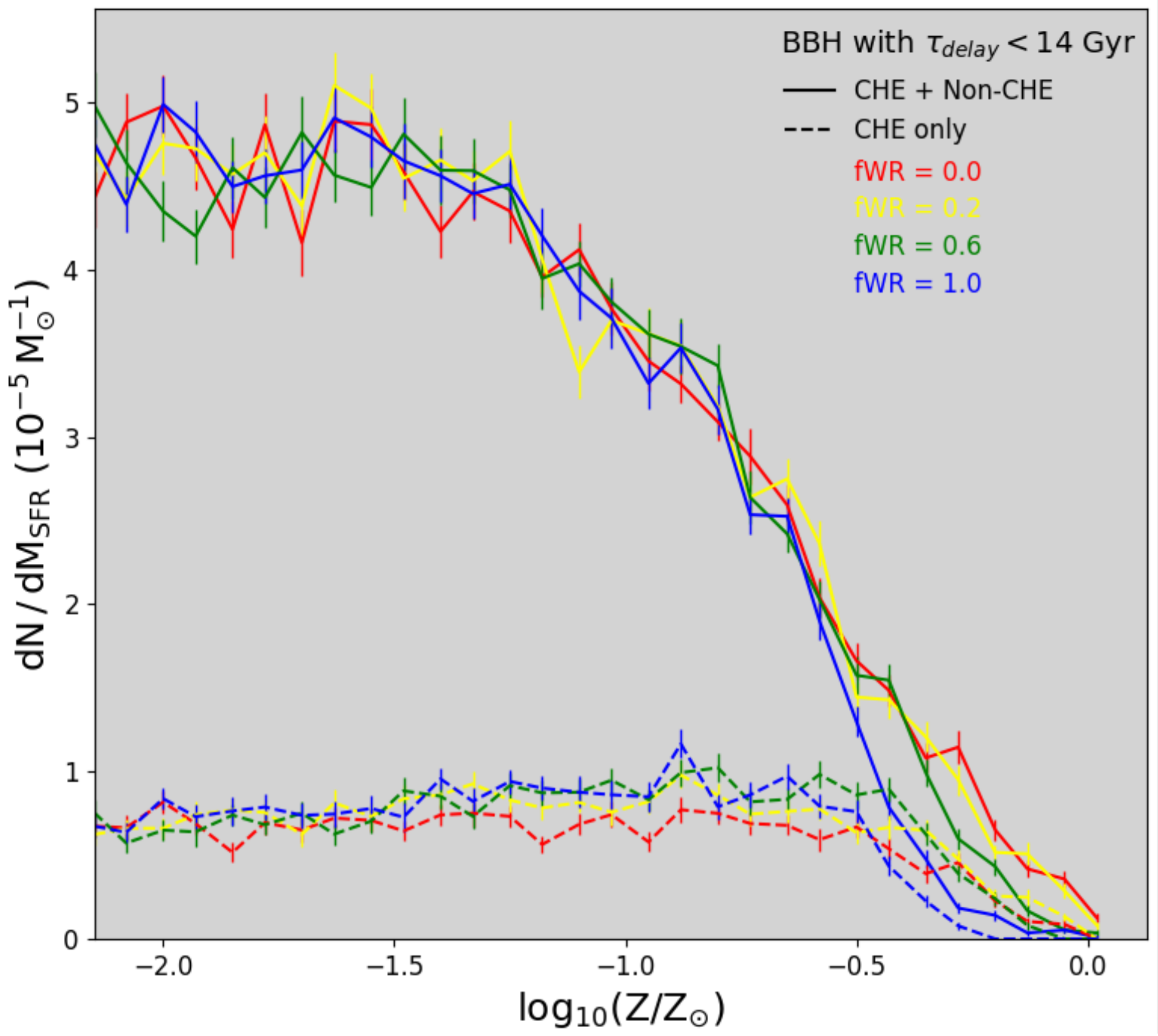}
    \caption{Yield of \acp{BBH} that will merge within 14 Gyr per unit star forming mass as a function of metallicity. The solid lines are the rates for the entire population --- both \ac{CHE} and non-\ac{CHE} binaries --- while the dashed lines are the rates for the \ac{CHE} binaries only. Colours indicate \ac{WR} mass loss rate multipliers. Error bars indicate 90\% confidence intervals from sampling uncertainty.
    }
    \label{fig:BBH_yield}
\end{figure}

Figure~\ref{fig:BBH_yield} shows the merging \ac{BBH} yield: the formation rate per unit star forming mass as a function of metallicity for systems that will merge within 14 Gyr. The solid lines are the rates for the entire population --- both \ac{CHE} and non-\ac{CHE} binaries --- while the dashed lines are the rates for \ac{CHE} binaries only. \ac{WR} mass loss multipliers are differentiated by the colour of the lines.

The overall yield of merging \acp{BBH} is quantitatively similar to the simulations of \citet{Neijssel_2019}, who predicted a yield of $\sim$ 6, 4, and 1 merging \acp{BBH} per $10^5\, M_\odot$ of star formation at $Z=0.01\, Z_\odot$, $0.1\, Z_\odot$, and $0.3\, Z_\odot$, respectively. The small differences are due partly to the inclusion of the \ac{CHE} channel as well as \acp{PISN} and \acp{PPISN} in this work, which were not included in \citet{Neijssel_2019}.  

Meanwhile, the low-metallicity \ac{CHE} channel yield of slightly less than 1 merging \acp{BBH} per $10^5\, M_\odot$ of star formation is similar to both the \citet{Mandel_2016} back-of-the-envelope estimate and the \citet{Marchant_2016} detailed models which indicate $\sim 0.7$ merging \acp{BBH} below the \ac{PISN} mass gap per 1000 core-collapse supernovae or per $10^5\, M_\odot$ of star formation at $Z=0.02\,Z_\odot$.  

The paucity of \ac{CHE} \acp{BBH} at high metallicity, \mbox{$Z \gtrsim 0.3\, Z_\odot$}, is due primarily to a combination of the upward shifting of the allowed initial periods at higher metallicities (see Figures \ref{fig:CHEBoundaries} and \ref{fig:pInitialVSmTotal}) and greater orbital widening by stronger high-metallicity winds.  The increase in orbital period at \ac{BBH} formation increases the delay times, preventing the \acp{BBH} from merging within 14 Gyr.   The widening by mass loss is ameliorated by reduced \ac{WR} mass loss rates.  However, the \ac{WR} mass loss multipliers have negligible effect at low metallicities because the total mass loss rate is too low even for $f_\mathrm{WR}=1$ (see Figure \ref{fig:WRActualMassLoss} and associated discussion).  \citet{Neijssel_2019} discuss the impact of metallicity on the non-\ac{CHE} \acp{BBH} yield, highlighting the contributions of wind-driven widening and stellar evolutionary stage at mass transfer.

\begin{figure}
    \vspace{0.5mm}
    \centering
    \includegraphics[viewport = 0 0 565 495, width=8.45cm, clip]{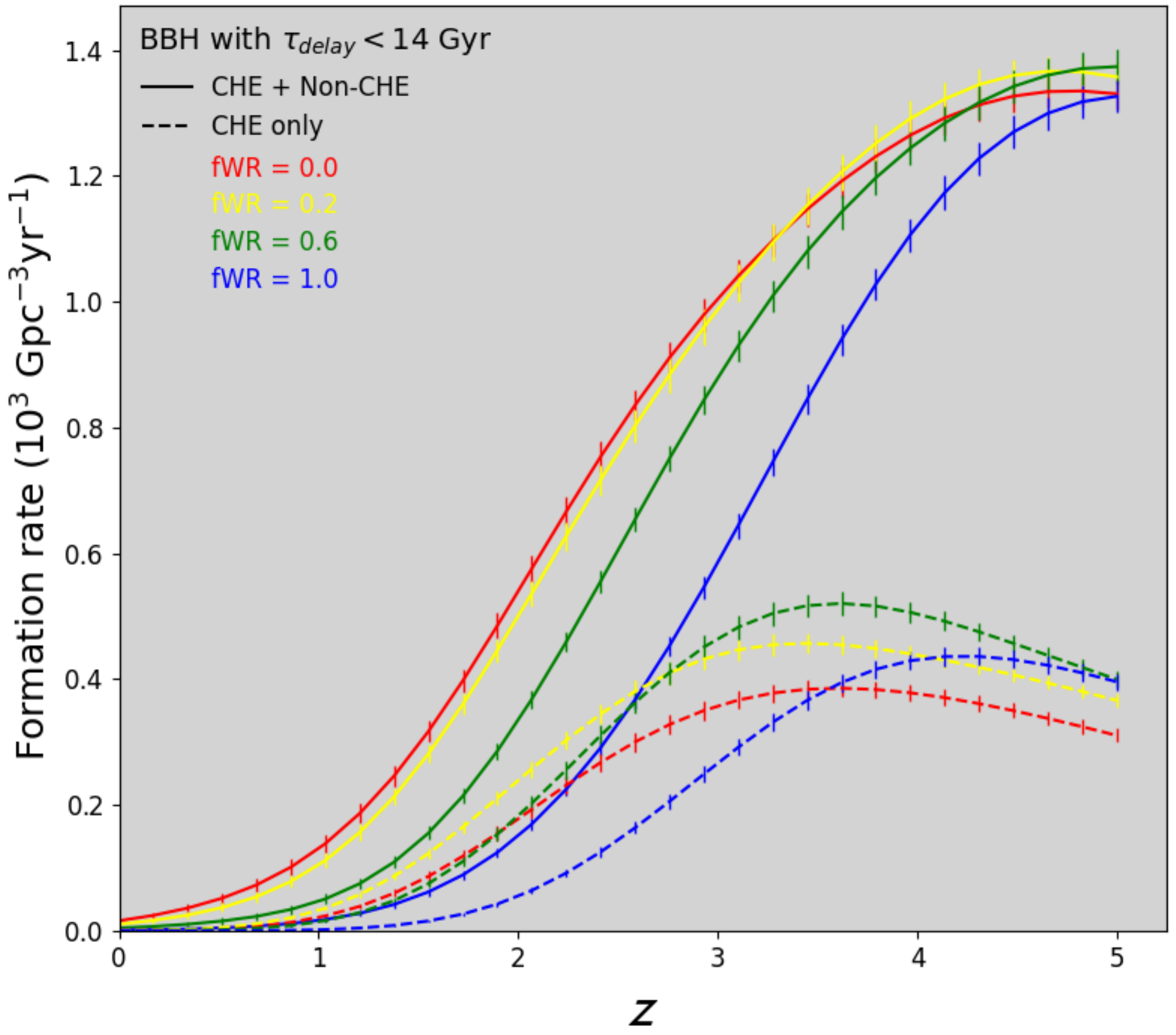}
    \caption{\ac{BBH} formation rate per \CubicGigaparsec\ of comoving volume per year as a function of redshift for \acp{BBH} that will merge within 14 Gyr. Error bars indicate sampling uncertainty.}
    \label{fig:redshift_formation}
\end{figure}

Figure~\ref{fig:redshift_formation} shows the \ac{BBH} formation rate per unit comoving volume per unit source time as a function of redshift. The formation rate for \ac{CHE} \acp{BBH} peaks at ${z\approx4.25}$ for ${f_\mathrm{WR}=1.0}$, and at ${z\approx3.5}$ for other \ac{WR} mass loss multipliers for the chosen metallicity-specific star formation rate history.  The \ac{BBH} formation rate for both \ac{CHE} and non-\ac{CHE} channels peaks at higher redshifts than the assumed star formation rate because both have higher yields per unit star formation at lower metallicities, which are prevalent at higher redshifts.

\subsubsection{Merger delay times}\label{subsubsec:results_merger_delays}

\begin{figure}
    \centering
    \includegraphics[viewport = 0 0 545 490, width=8.45cm, clip]{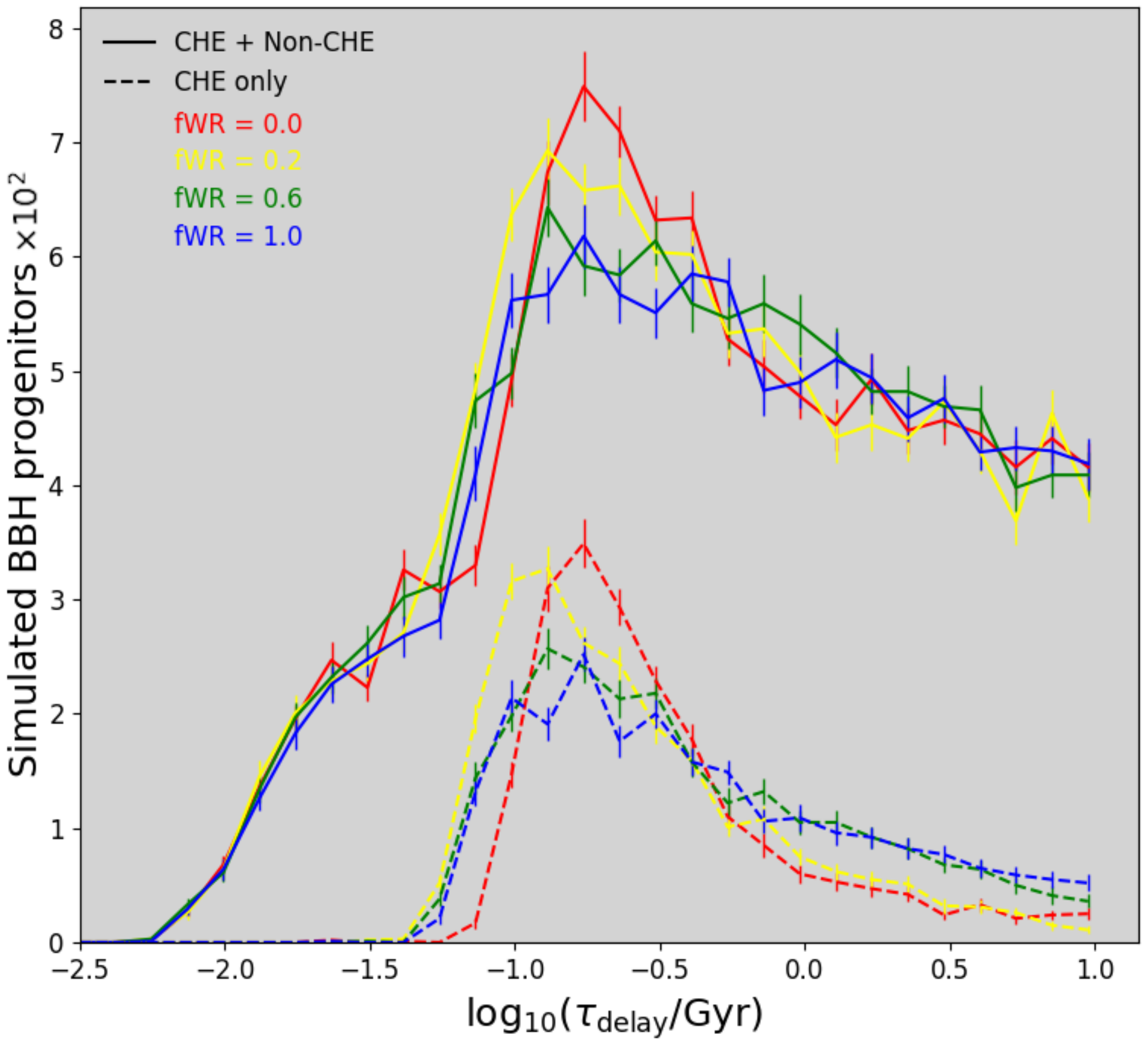}
    \caption{Distribution of delay times between formation and merger for \acp{BBH}.  All metallicities from the simulation are combined with equal weights and arbitrary counts per uniform bins in log delay time are shown. Error bars indicate sampling uncertainty. 
    }
    \label{fig:delayTimes}
\end{figure}

Figure~\ref{fig:delayTimes} indicates the distribution of delay times between star formation and \ac{BBH} mergers.  This figure combines all metallicities with equal weights, without considering their contribution to the observable systems, so should be viewed as an indicative sketch.  

Non-\ac{CHE} binaries in Figure~\ref{fig:delayTimes} have a very broad distribution of delay times.  Some are very short, less than 10 Myr, due to significant hardening during mass transfer episodes, including through dynamically unstable mass transfer and common-envelope ejection, as well as fortuitously directed supernova natal kicks.  Meanwhile, there is an almost flat tail of long delay times on this logarithmic plot, corresponding to a $p(\tau_\mathrm{delay}) \sim 1/\tau_\mathrm{delay}$ distribution.

On the other hand, binaries formed through \ac{CHE} are seen to have a more strongly clustered delay time distribution, with typical delay times of between 100~Myr and 1~Gyr.  There are no ultra-short delay times because, with the exception of \ac{RLOF} at \ac{ZAMS}, such binaries do not undergo mass transfer that could harden the binary.  Moreover, the high masses of \ac{CHE} stars imply that they do not experience asymmetric supernovae and associated natal kicks in the \COMPAS model.

The smallest time delay between formation and merger for \ac{CHE} systems in our simulations ranges from \mbox{$\sim0.025$\,Gyr}\ for $f_\mathrm{WR}=0.0$ to \mbox{$\sim0.033$\,Gyr}\ for $f_\mathrm{WR}=1.0$.  The combination of lower metallicities and reduced mass loss rates yields the shortest delay times, allowing binaries to start evolution from closer separations while avoiding L2 overflow and to avoid subsequent widening through mass loss.  This is consistent with the minimal delay times found in other studies.  \citet{Mandel_2016}, who consider only $Z=0.004\,Z_\odot$, estimate minimum delay times of \mbox{$\sim3.5$\,Gyr}. \citet{Marchant_2016} find minimal delay times of \mbox{$\sim0.4$\,Gyr} and point out the metallicity dependence. \citet{duBuisson_2020} consider the lowest metallicities among these studies, $Z=10^{-5}$, and find the shortest delay times, \mbox{$\sim0.02$\,Gyr}.

Some \ac{CHE} binaries will be significantly widened by mass loss, potentially losing up to a factor of $\sim 2$ in mass during the \ac{WR} phase (see Figure \ref{fig:WRActualMassLoss}) and thereby increasing their separation by the same factor.  The gravitational-wave driven coalescence time scales as $a^4 M^{-3}$ \citep{Peters_1964}, so a factor of 2 each in mass decrease and semi-major axis increase would yield a factor of $2^7 \sim 100$ increase in the delay time.  This explains the long delay time tail of the \ac{CHE} \ac{BBH} distribution, as well as the decrease in the prominence of this tail as the \ac{WR} wind mass loss multiplier is reduced.  Even when $f_\mathrm{WR}=0$, some \ac{CHE} \acp{BBH} will have long delay times due to the mass lost in \acp{PPISN}.  

\subsubsection{Merger rates}
\label{subsubsec:results_merger_rates}

\begin{figure}
    \centering
    \includegraphics[viewport = 0 0 550 495, width=8.45cm, clip]{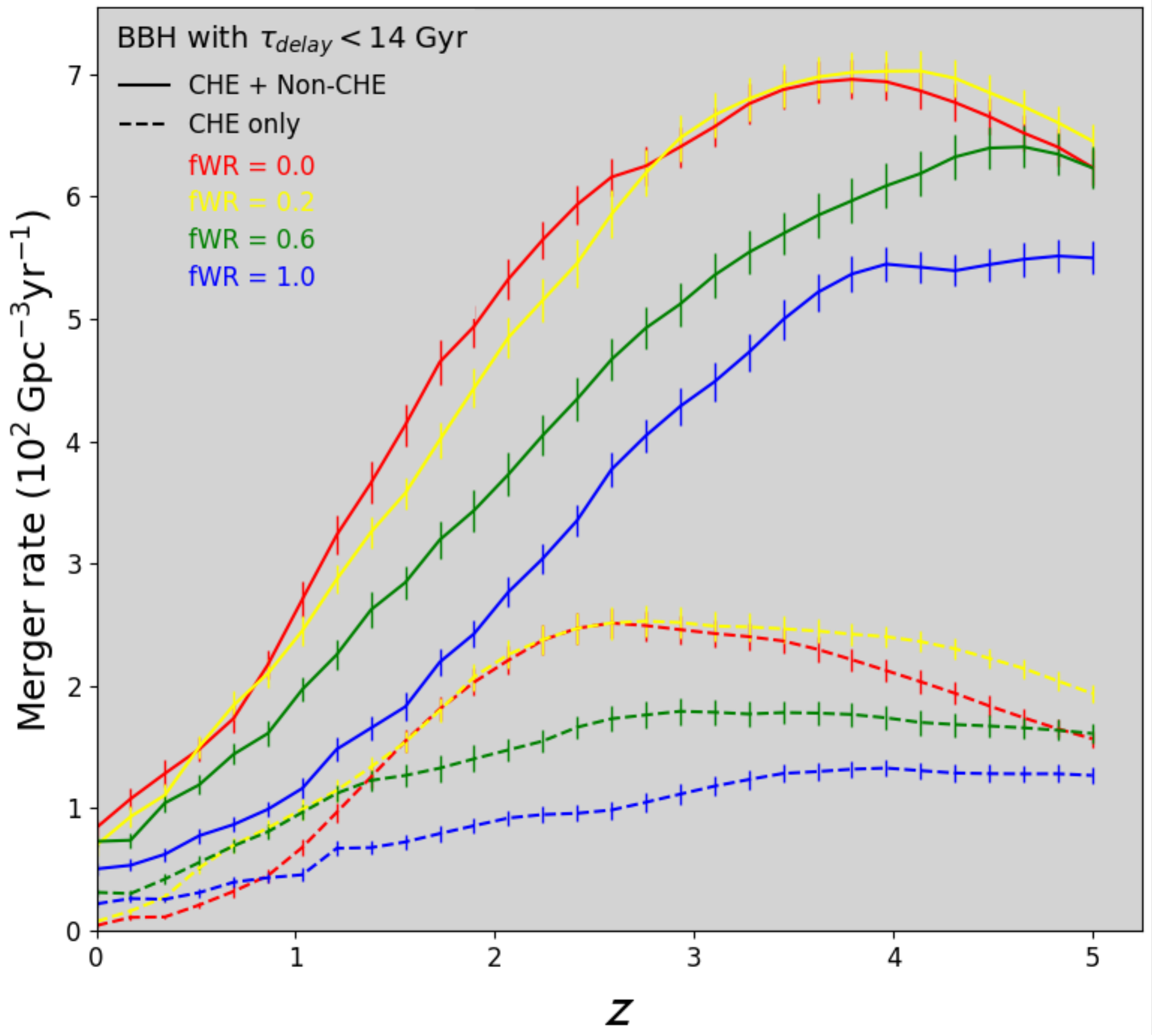}
    \caption{\ac{BBH} merger rate per \CubicGigaparsec\ of comoving volume per year of source time as a function of redshift. Error bars indicate sampling uncertainty.
    \label{fig:redshift_merger}
    }
\end{figure}

Figure~\ref{fig:redshift_merger} shows the \ac{BBH} merger rate per \CubicGigaparsec\ of comoving volume per year of source time as a function of redshift. The merger rate for \ac{CHE} \acp{BBH} peaks at $z\approx4$ for $f_\mathrm{WR}=1.0$, and at $z\approx3$ for other \ac{WR} mass loss multipliers. Both \ac{CHE} and total \ac{BBH} merger rates peak at higher redshifts than the star formation rate, which peaks at $z\approx 2$ (see Figure~\ref{fig:madau_plot}), because both \ac{CHE} and non-\ac{CHE} channels have higher yields at lower metallicity (see Figure~\ref{fig:BBH_yield}).  
The relatively small difference between the peak formation and merger rates is explained by the short delay times for the \ac{CHE} systems (see Figures~\ref{fig:redshift_formation} and \ref{fig:delayTimes}).  The delay times are particularly short for $f_\mathrm{WR}=0$ \ac{CHE} \acp{BBH}, which explains their suppressed merger rate in the local Universe.
  
\begin{figure}
    \centering
    \includegraphics[viewport = 5 5 570 495, width=8.45cm, clip]{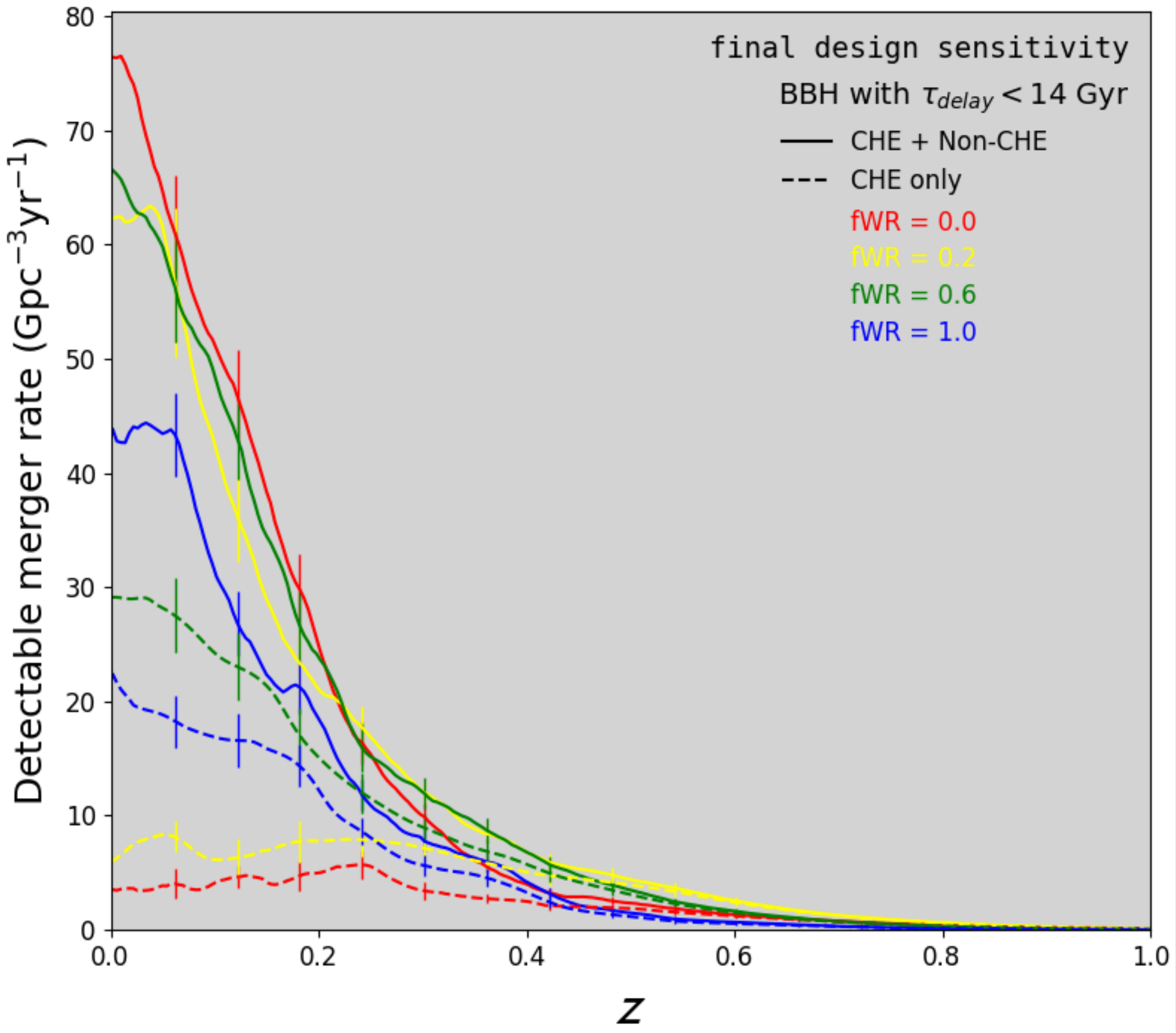}
    \caption{The merger rate of \acp{BBH} detectable by \ac{aLIGO} at final design sensitivity, as a function of merger redshift. Error bars indicate sampling uncertainty.}
    \label{fig:redshift_merger_observable}
\end{figure}

The merger rates of \acp{BBH} that could be observed by \ac{aLIGO} operating at final design sensitivity merger rates are shown in Figure~\ref{fig:redshift_merger_observable}. Binaries formed through \ac{CHE} have higher average masses than non-\ac{CHE} binaries, which increases the range within which they are detectable by \ac{aLIGO}.  Therefore, \ac{CHE} \acp{BBH} make up a higher fraction of all detections at greater redshifts.

\subsubsection{aLIGO detection rates}
\label{subsubsec:results_detection_rates}

\begin{figure}
    \centering
    \includegraphics[viewport = 5 5 565 495, width=8.45cm, clip]{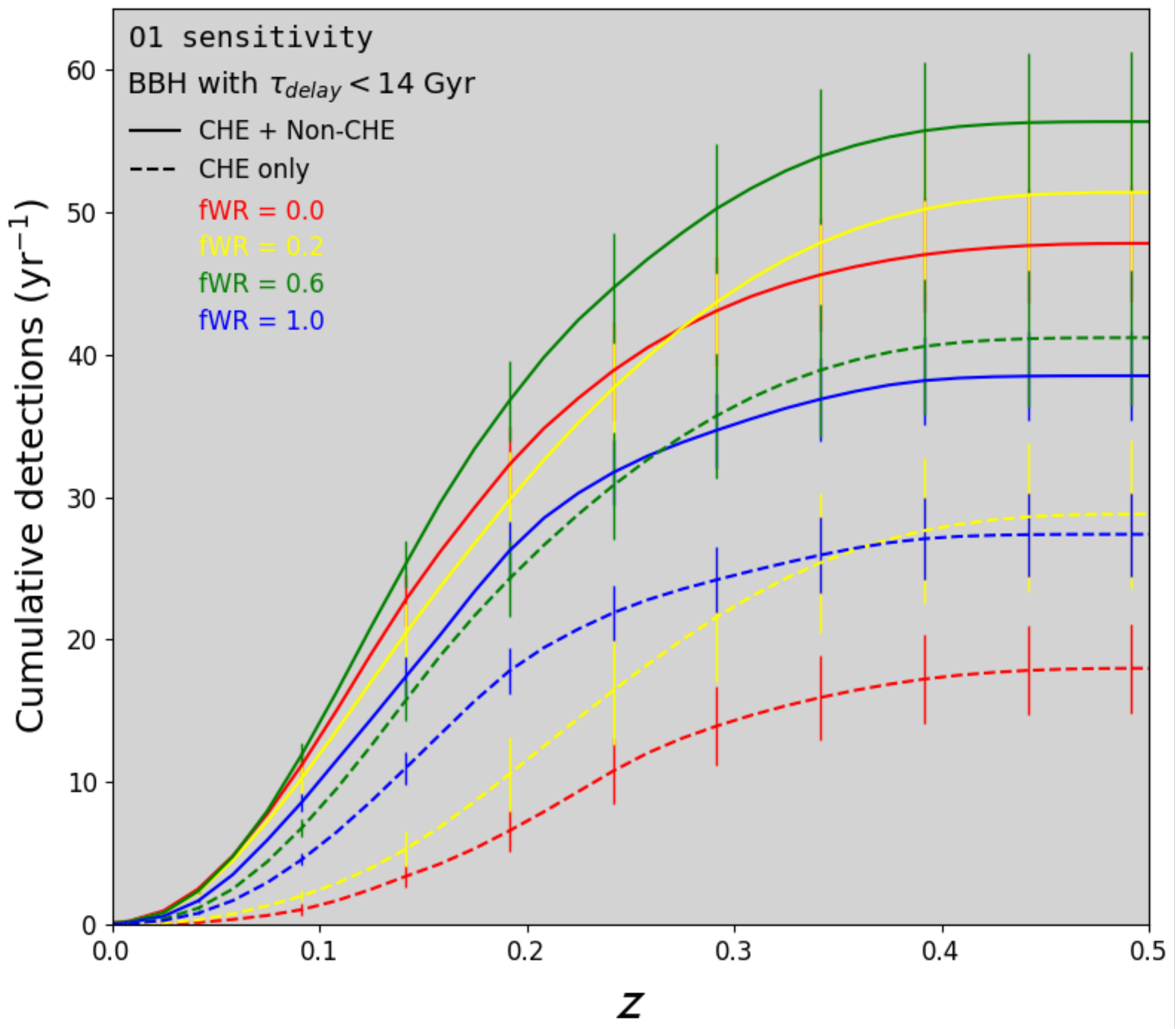}
    \caption{Cumulative \ac{BBH} detections as a function of merger redshift, per year of observing at \ac{aLIGO} O1 sensitivity. Error bars indicate sampling uncertainty.}
    \label{fig:redshift_detection_o1}
\end{figure}

\begin{figure}
    \centering
    \includegraphics[viewport = 5 5 565 495, width=8.45cm, clip]{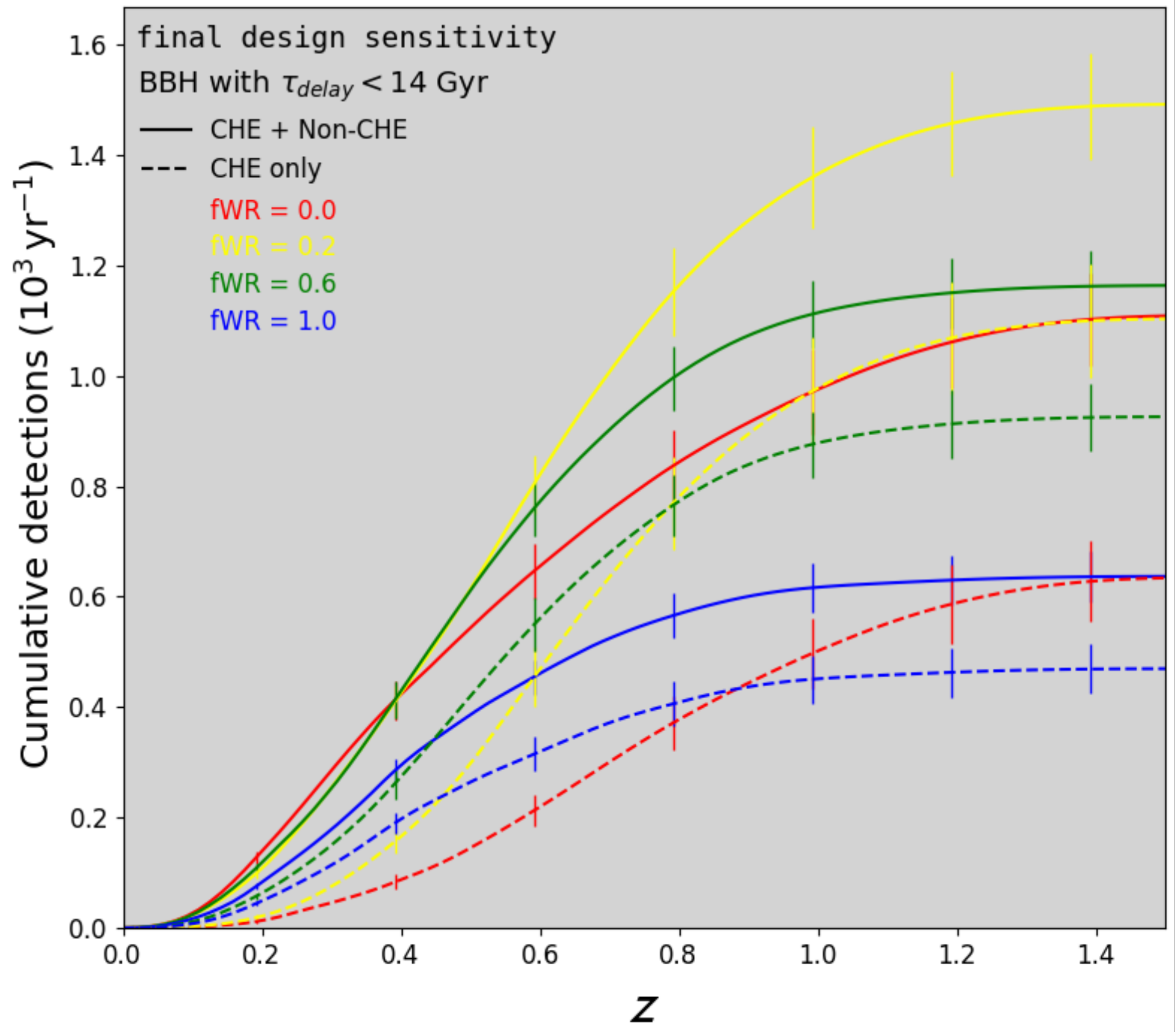}
    \caption{Cumulative \ac{BBH} detections as a function of merger redshift, per year of observing at \ac{aLIGO} final design sensitivity. Error bars indicate sampling uncertainty.}
    \label{fig:redshift_detection_design}
\end{figure}

Figures~\ref{fig:redshift_detection_o1} and \ref{fig:redshift_detection_design} show the predicted cumulative detection rates per year of observing time as a function of redshift for \ac{aLIGO} O1 and final design sensitivities, respectively. 

Figure \ref{fig:redshift_detection_o1} shows that the total expected detection rate at O1 sensitivity is 38--55 detections per year, depending on the assumed value of the \ac{WR} mass loss rate multiplier.    This would correspond to 17--25 detections over the 166 days of coincident data over the first two advanced detector observing runs, O1 and O2.  In fact, only 10 \acp{BBH} were observed during this time \citep{BBH:O1O2}.    

The increased detection rate relative to the preferred metallicity-specific star formation rate model of \citet{Neijssel_2019}, who predicted 22 detections per year, in agreement with the O1 and O2 observations, is due to the contribution of \ac{CHE} \acp{BBH}.   \ac{CHE} \acp{BBH} may constitute up to $\sim 70\%$ of all \ac{BBH} detections at both the O1 sensitivity and the final design sensitivity of \ac{aLIGO}.  

The star formation history model of \citet{Neijssel_2019} was tuned to the gravitational-wave observations, and explaining the relatively high masses of observed \acp{BBH} required significant high-redshift, low-metallicity star formation.  The inclusion of \ac{CHE} \acp{BBH} naturally yields a population of high-mass sources, allowing for the high-mass star formation rate to be reduced in line with the \citet{Madau_2014,Madau_2017} models.  This would naturally bring rate predictions in line with the O1 and O2 observations and correspondingly reduce predicted detection rates for future detectors.

Using the preferred cosmic metallicity star formation model of \citet{Neijssel_2019}, as we do here, and assuming $f_\mathrm{WR}=1$, we predict a total \ac{BBH} detection rate of $\approx 660$ per year at \ac{aLIGO} design sensitivity (vs.~ $\approx 37$ at O1 sensitivity), with $\approx 470$ ($\approx 27$) of these coming from the \ac{CHE} channel.   The \ac{CHE} \ac{BBH} detection rates are a factor $\sim 2$ larger than those estimated by \citet{duBuisson_2020}, who found that $\approx 250$ ($\approx 13$) \ac{CHE} \acp{BBH} per year may be detected at \ac{aLIGO} design (O1) sensitivity.  The differences in the assumed metallicity-specific star formation rates in these studies are responsible for much of this difference.

We note that in both Figures \ref{fig:redshift_detection_o1} and \ref{fig:redshift_detection_design} the order of the lines with respect to the number of detections does not match the order of the \ac{WR} mass loss multipliers.  This is due to the interplay between the formation rate of \acp{BBH} and their delay times as a function of $f_\mathrm{WR}$, which are described in Figures~\ref{fig:redshift_formation} and~\ref{fig:delayTimes} and associated discussion.  For example, in the absence of \ac{WR} winds ($f_\mathrm{WR}=0$), reduced delay times due to a lack of binary widening relative to simulations with higher \ac{WR} mass loss rates mean that very few \ac{CHE} \acp{BBH}, which predominantly form at lower metallicities and thus higher redshifts, merge in the local Universe, where they would be detectable.

\subsubsection{Mass distribution of detectable \ac{BBH} mergers}
\label{subsubsec:results_chirp_mass}

\begin{figure}
    \centering
    \includegraphics[width=8.45cm]{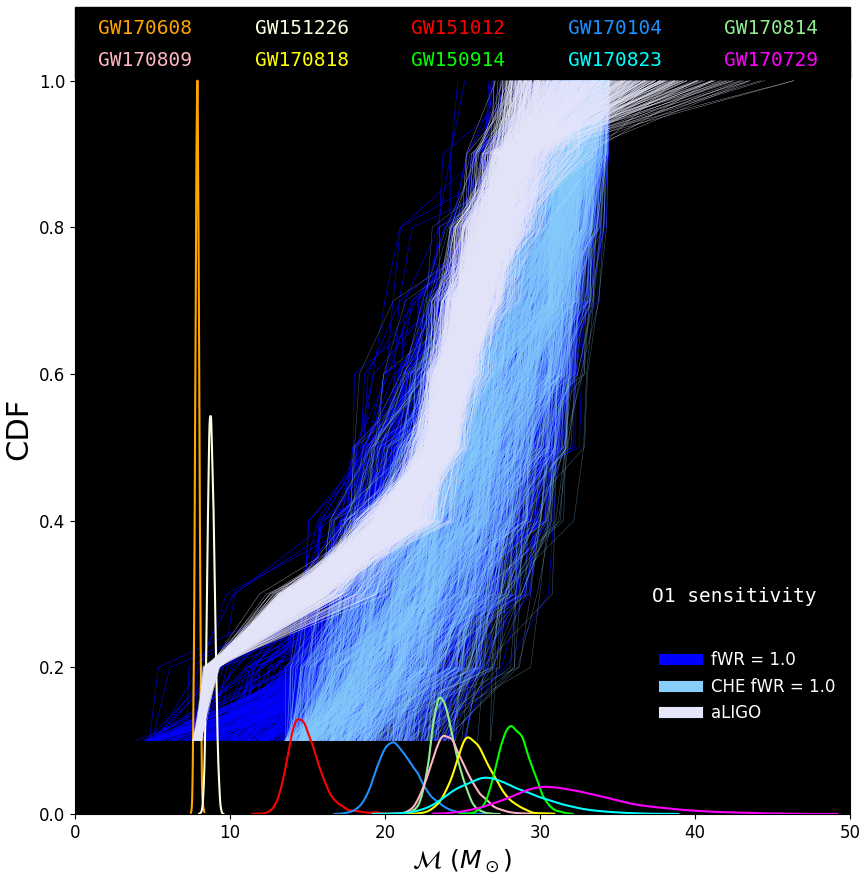}
    \caption{Chirp mass posteriors for the 10 \ac{BBH} mergers detected during the first and second \ac{aLIGO} observing runs \citep{Abbott_2019} are shown in colour at the bottom, with labels at top.  These are randomly sampled to construct the cumulative density functions shown in lavender (each curve corresponds to a cumulative distribution through 10 samples, one from each posterior).  Cumulative density functions for \COMPAS chirp mass predictions based on $f_\mathrm{WR}=1.0$ models are also based on 10 samples from either the full population (dark blue lines) or \ac{CHE} systems only (light blue lines).}
    \label{fig:cumulativeMchirpLigo}
\end{figure}

The cumulative distribution functions for the modelled chirp mass distribution of detectable \ac{BBH} mergers are shown in Figure~\ref{fig:cumulativeMchirpLigo}.  The dark blue lines indicate the chirp mass distribution of all \acp{BBH} while the light blue lines indicate the chirp mass distribution of \ac{CHE} \acp{BBH}.  In both cases, results for the \ac{WR} mass loss multiplier $f_\mathrm{WR}=1.0$ are reported, and the O1 \ac{aLIGO} sensitivity, which is similar to that of the second observing run, is used.  We show cumulative distribution functions for sets of 10 randomly selected samples from the \COMPAS models -- the number of \acp{BBH} detected during O1 and O2 -- in order to indicate the variation due to sampling fluctuations.  To avoid granularity due to the discreteness of the metallicity grid in \COMPAS models \citep[see][for a discussion]{Dominik:2015,Neijssel_2019}, we used continuous sampling in metallicity to construct the model predictions for this plot.

\begin{figure}
    \centering
    \includegraphics[viewport = 0 0 565 490, width=8.45cm, clip]{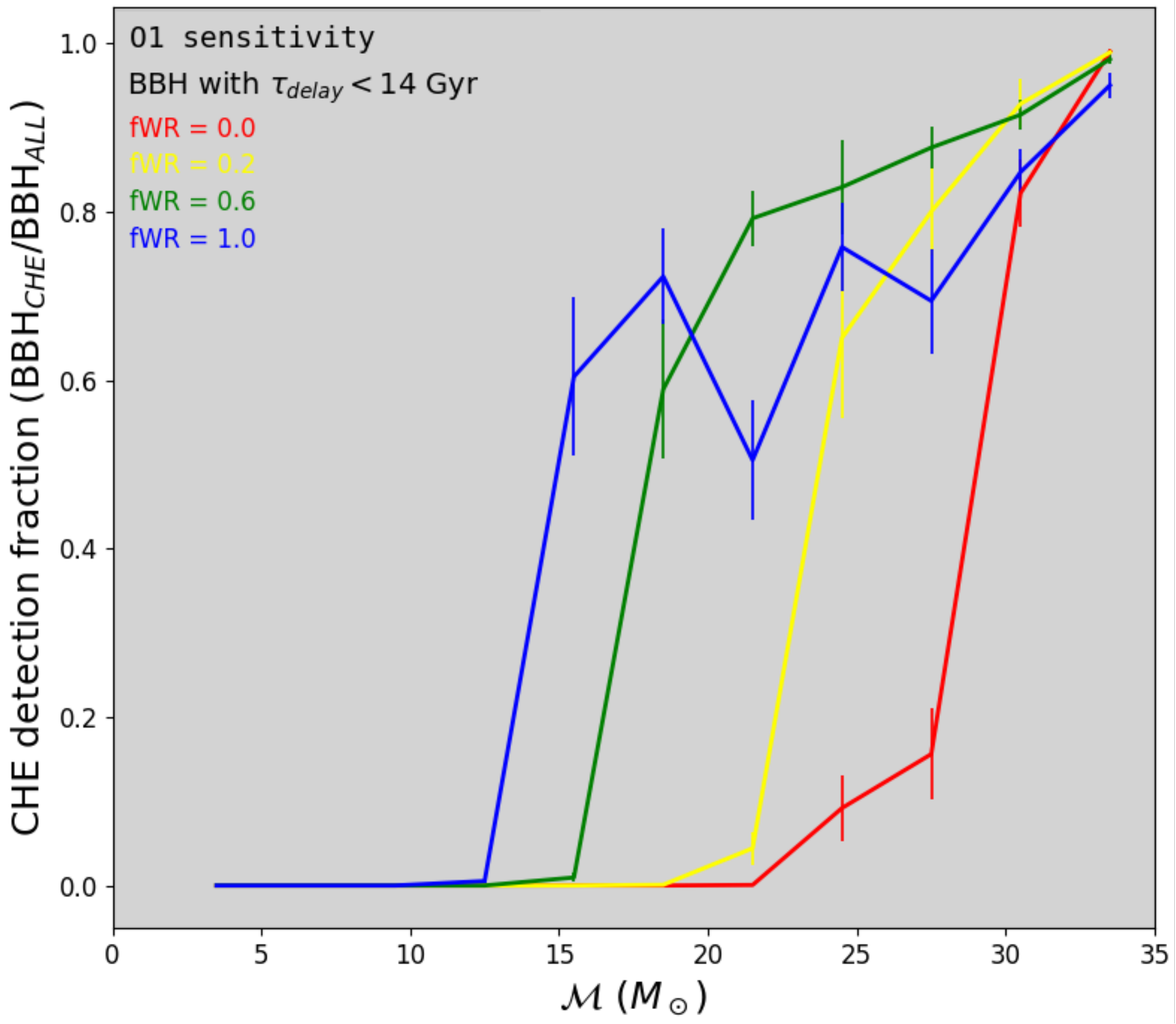}
    \caption{The fraction of \acp{BBH} formed through the \ac{CHE} channel among all \acp{BBH} detectable at \ac{aLIGO} O1 sensitivity, plotted as a function of chirp mass.}
    \label{fig:MchirpCHEfrac_O1}
\end{figure}

As mentioned previously, \ac{CHE} \acp{BBH} are more massive than typical non-\ac{CHE} \acp{BBH}.  The initial masses of \ac{CHE} \acp{BBH} must be high to allow for \ac{CHE} (see Figure \ref{fig:CHEBoundaries}).  Moreover, \ac{CHE} in our model allows stars to convert all of their mass to helium, whereas non-\ac{CH} massive stars typically have $\gtrsim 50\%$ of their mass in hydrogen-rich envelopes, which they lose prior to collapse into black holes in the course of binary evolution.  This is highlighted in Figure~\ref{fig:MchirpCHEfrac_O1}, which indicates the fraction of all \acp{BBH} detectable at \ac{aLIGO} O1 sensitivity that formed through the \ac{CHE} channel, as a function of chirp mass.  The \ac{CHE} channel dominates the production of \acp{BBH} at high chirp masses, particularly for reduced \ac{WR} mass loss models, when it yields increasingly large chirp masses ($\gtrsim 30\, M_\odot$ in the absence of \ac{WR} winds).

Figure~\ref{fig:cumulativeMchirpLigo} allows for a direct comparison between the modelled chirp mass distribution and the \ac{aLIGO} observations from the first two observing runs.  The individual posterior samples from the 10 \ac{aLIGO} \ac{BBH} detections during those observing runs are plotted at the bottom of the plot.   Randomly sampled cumulative distribution functions of the chirp mass of observed events are constructed by taking 10 random samples, one from each of the 10 \ac{aLIGO} observation posteriors and displayed as light lavender curves.  The overlap of the lavender and dark blue lines in Figure~\ref{fig:cumulativeMchirpLigo} shows that the \COMPAS model of \ac{BBH} formation, which includes the contribution of \ac{CHE}, yields a chirp mass distribution of detectable \ac{BBH} mergers that is consistent with detections during the first two \ac{aLIGO} observing runs.

\section{Concluding remarks}\label{sec:conclusion}

We described the model of \acf{CHE} that we implemented in the rapid population synthesis code \COMPAS. We used MESA models to determine the critical rotation thresholds for \ac{CHE}, and provided fits that can be used in other rapid binary population synthesis applications.   We synthesised 12 million binary systems over a range of metallicities (30 metallicities evenly spaced across the range \mbox{$\minus4\leq\logten{Z}\leq\minus1.825$}) and \ac{WR} wind mass loss multipliers (\mbox{$f_\mathrm{WR}\in\{\,0.0\,, 0.2\,, 0.6\,, 1.0\,\}$}).  We confirmed that our simplified models match detailed binary evolution simulations \citep{Marchant_2016,duBuisson_2020} well.  

We investigated the contribution of \ac{CHE} and non-\ac{CHE} channels to \ac{BBH} formation under a single set of assumptions.  We found that the \ac{CHE} channel may contribute more than half, and perhaps as much as three quarters, of all \ac{aLIGO} \ac{BBH} detections arising from isolated binary evolution.  \ac{CHE} \acp{BBH} may represent $\gtrsim 80\%$ of detectable sources with the highest chirp masses of $\gtrsim 30\,M_\odot$.  A comparison between our model population and the population of detected binaries from the first two advanced detector observing runs indicates that the current model over-predicts the total number of sources by a factor of $\sim 2$, but matches the observed chirp mass distribution.

We made a number of simplifying assumptions in this study that can be investigated and improved on in the future.  We generally erred on the side of being conservative about \ac{CHE} predictions:  
\begin{itemize}
\item We used \citet{Hurley_2000} non-rotating \ac{MS} models to set the radii and mass loss rates of \ac{CH} \ac{MS} stars.  The imperfect radius model for rapidly-rotating \ac{CH} stars in turn leads to differences in the orbital separation boundary for avoiding merger through L2 overflow between the \ac{COMPAS} and MESA models (see Appendix~\ref{sec:che_thresholds}).
\item We used simplified tidal interaction assumptions under which \ac{CH} stars are immediately tidally synchronised, yet do not store angular momentum.  Accounting for the angular momentum stored in stars -- and the additional angular momentum carried away by winds from a rotating star -- impacts the response of the binary's orbit to mass loss, and reduces the amount of orbital widening by wind mass loss in close binaries.
\item Contrary to our simplified assumptions, winds may interact with binary companions.  This is particularly true in close binaries, when the wind speeds are comparable to the orbital speeds, and wind interactions may produce additional drag and reduce the amount of orbital widening \citep[e.g.,][]{BrookshawTavani:1993,MacLeodLoeb:2020}.
\item We ignored the possibility of initially non-\ac{CH} stars switching to \ac{CHE} in response to mass accretion.   
\item We assumed that all mass loss in \acp{PPISN} happens instantaneously, rather than over several pulsations (although the first pulsation is likely to be dominant, so this approximation may not be especially problematic).  
\end{itemize}

Our predicted \ac{BBH} merger rate at redshift zero of 50\PerCubicGigaparsecPerYear\ (including 20\PerCubicGigaparsecPerYear\ from the \ac{CHE} channel) for the default \ac{WR} mass loss rate $f_\mathrm{WR}=1.0$ over-estimates the number of \ac{BBH} detections during the first two observing runs of gravitational-wave detectors.  This is at least partly due to our using a metallicity-specific star formation rate prescription from \citet{Neijssel_2019} that was designed to reproduce gravitational-wave observations without accounting for \ac{CHE}.  A resolution may involve reducing the high-redshift star formation rate back to levels more closely matching the models of \citet{Madau_2014,Madau_2017}.  

Two other observational constraints on \ac{CHE} \ac{BBH} formation come from the spins of observed \ac{BBH} mergers and from potential electromagnetic observations of their progenitors.   At first glance, the effective spins of \acp{BBH} observed to date \citep{BBH:O1O2} do not match the large reservoirs of angular momentum in \ac{CH} stars.   However, \ac{WR} winds can carry away much of angular momentum.  \citet{Marchant_2016} argued that typical dimensionless effective spins of \ac{CHE} sources should be $\sim 0.4$, much lower than the super-critical spins expected at \ac{WR} star formation.  The fraction of stellar angular momentum lost in winds during the \ac{WR} phase can be estimated as 
\begin{equation}
\frac{\Delta L}{L} \sim \frac{2}{3} \left(\frac{R_\mathrm{WR}}{R_\mathrm{WR,g}}\right)^2 \frac{\Delta M}{M},
\end{equation}
where the ratio of the radius of the \ac{WR} star to its gyration radius is $R_\mathrm{WR}/R_\mathrm{WR,g} \sim 10$.  Thus, Wolf-Rayet winds could lose the overwhelming bulk of the angular momentum that \ac{CHE} stars have, as long as $\Delta M / M > 0.01$, which is true even at $Z=0.01\,Z_\odot$ if \ac{WR} mass loss is not suppressed (see Figure \ref{fig:WRActualMassLoss}).  Unlike binaries that are hardened during the common-envelope phase to the point where tides can efficiently spin up the WR companion \citep{Kushnir:2016,Belczynski:2020,Bavera:2020}, binaries evolving through the \ac{CHE} channel will tidally decouple during the \ac{WR} phase.  For example, a typical binary from our simulations with component \ac{ZAMS} masses of 60 M$_\odot$ at metallicity $Z=0.00089$ and an initial orbital period of just over 1 day will evolve through the \ac{CHE} channel and form a merging \ac{BBH} with individual black hole masses of 38 M$_\odot$.  By the end of the \ac{WR} phase, following mass loss with $f_\mathrm{WR}=1.0$, the components will have masses of 52 M$_\odot$ and an orbital separation of 25 R$_\odot$.  The \ac{WR} radius at this time is $<2 \mathrm{R}_\odot$ \citep[e.g.,][]{Yoon:2012}, so the tidal synchronisation timescale will be several hundred Myr.  This is much longer than the duration of the \ac{WR} phase, so once spun down by winds, these stars cannot be spun up again by tides.

\ac{CHE} \ac{BBH} progenitors could yield interesting observational candidates.  Systems such as WR20a \citep{Rauw:2004} and BAT99-32 \citep{Shenar:2019} may belong in this category.  The metallicity of the Galaxy is too high to allow for merging \ac{CHE} \acp{BBH} according to our models, but we expect them to be formed at a rate of $\sim 3 \times 10^{-6}$ per year in the Magellanic clouds.  Given the typical \ac{MS} and \ac{WR} phase lifetimes of $3\times 10^6$ and $3\times 10^5$ years, respectively, we may hope to detect $\sim 10$ \ac{MS} \ac{CH} binaries and $\sim 1$ binary composed of two naked helium stars formed through \ac{CHE} and en route to \ac{BBH} formation in the Magellanic clouds today.

The joint model for the classical and \ac{CHE} isolated binary evolution channels developed here will enable simultaneous inference on binary evolution model parameters and the metallicity-specific star formation history once the full trove of observations from the third gravitational-wave observing run is available.  Ultimately, the relatively short delay times of \ac{CHE} \acp{BBH} make them ideal probes of high-redshift star formation history, while their high masses make them perfect targets for third-generation gravitational-wave detectors with good low-frequency sensitivity, such as the Einstein Telescope \citep{ET} or the Cosmic Explorer \citep{CE}.

\section*{Acknowledgements}\label{sec:ack}
Simulations in this paper made use of the \COMPAS rapid population synthesis code, which is freely available at \url{http://github.com/TeamCOMPAS/COMPAS}.  
The version of \COMPAS used for these simulations was v02.11.01a, built specifically for these simulations; functionality in this release was integrated into the public \COMPAS code base in v02.11.04. 

The authors thank Selma de Mink and other colleagues in Team COMPAS, as well as Morgan MacLeod, for helpful discussions. 
We also thank Tim Riley for assistance in running COMPAS simulations.  
IM is a recipient of the Australian Research Council Future Fellowship FT190100574.  
AVG acknowledges funding support by the Danish National Research Foundation (DNRF132).

\section*{Data availability}
The data underlying this article will be available via \url{https://zenodo.org/communities/compas/}.

\bibliographystyle{aasjournal}
\bibliography{bib.bib}


\appendix
\bigskip
\section{CHE thresholds}\label{sec:che_thresholds}

We evolved single stars over a range of masses, metallicities, and rotational frequencies with version 10108 of the \ac{MESA} code in order to find the boundary between \ac{CHE} and regular non-\ac{CH} stellar evolution.\footnote{The complete set of MESA input files necessary to reproduce these simulations will be made available after acceptance of the manuscript.}  Simulations were performed until the end of the main sequence without mass loss, while enforcing solid body rotation at a constant angular frequency.

Opacities are computed using tables from the OPAL project \citep{IglesiasRogers1996} with solar-scaled metal mass fractions as given by \citet{GrevesseSauval1998}. The equation of state is a combination of the OPAL \citep{RogersNayfonov2002}, HELM \citep{TimmesSwesty2000}, PC \citep{PotekhinChabrier2010} and SCVH \citep{Saumon1995} equations of state. Nuclear reaction rates are taken from \citet{CaughlanFowler1988} and \citet{Angulo+1999} with preference for the latter when available.

Our choices for overshooting and rotational mixing processes follow those of \citet{Brott+2011}. Namely, overshooting from convective hydrogen burning cores is modeled as step overshooting, increasing the size of the convective core by $0.335 H_P$, where $H_P$ is the pressure scale height at the edge of the convective boundary. As we consider solid body rotation, the only significant mixing process included in our simulations is the effect of Eddington-Sweet circulations as described by \citet{Kippenhahn1974}, with an efficiency factor of $1/30$ \citep{ChaboyerZahn1992,Heger+2000}. We also include the inhibiting effect of composition gradients in rotational mixing as described by \citet{Heger+2000}, given by the dimensionless parameter $f_\mu=0.1$ \citep{Yoon+2006}. The star was considered to evolve chemically homogeneously if the difference between the helium fraction across the star did not exceed $0.2$.

\begin{figure*}
\begin{center}
\includegraphics[viewport = 0 0 1200 625, width = 17.725cm, clip]{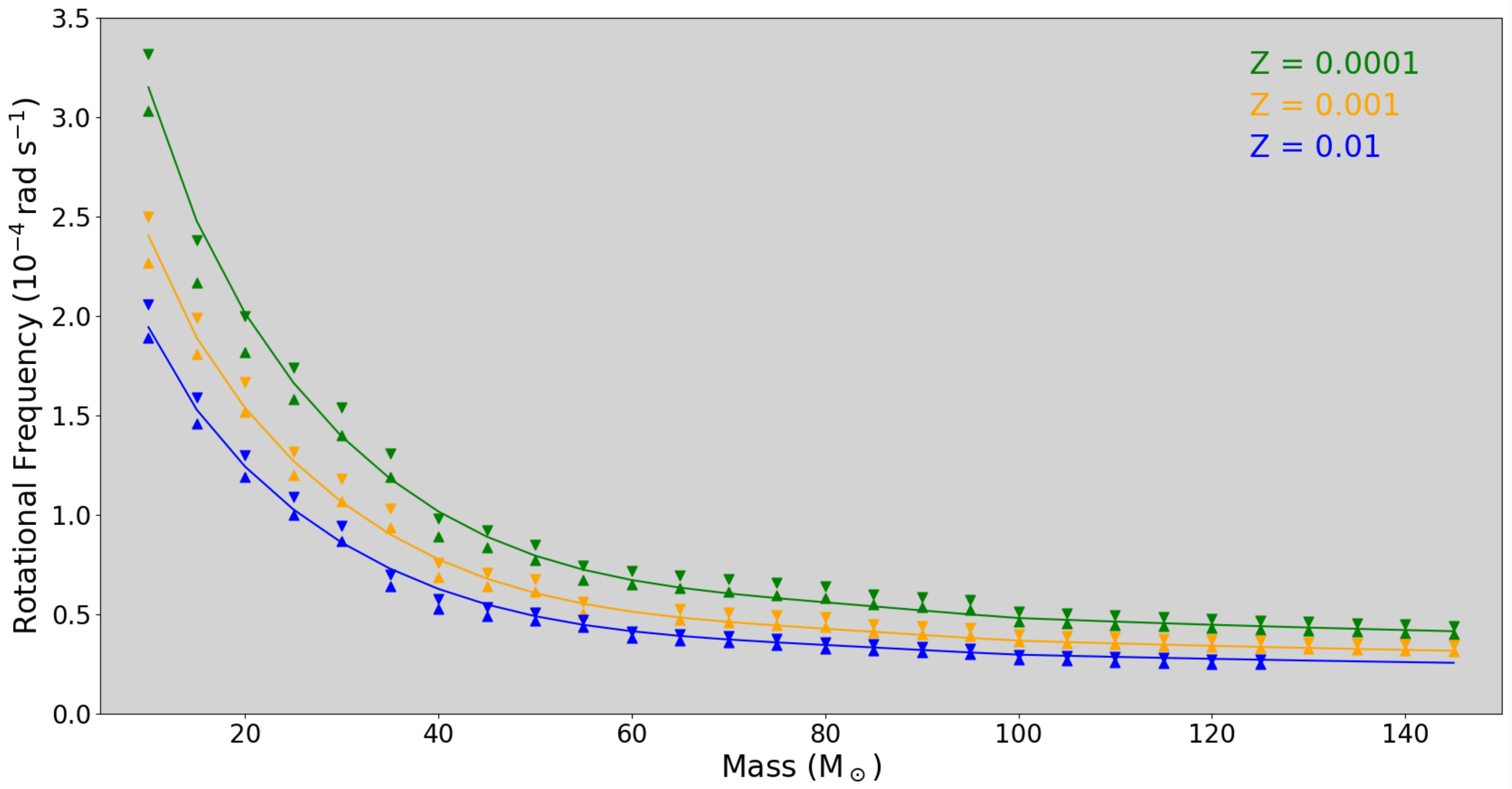}
\end{center}
\vspace{-10pt}
\captionof{figure}{Rotational frequency threshold for chemically homogeneous evolution as a function of mass and metallicity.  Downward and upward triangles represent the slowest-rotating \ac{CHE} model and fastest-rotating non-\ac{CHE} model at the given mass and metallicity, respectively.  The curves indicate the fits of Eqs.~(\ref{eq:CHEthreshold1}) and (\ref{eq:CHEthreshold2}).}
\label{fig:CHEfits}
\end{figure*}

Figure~\ref{fig:CHEfits} shows the maximum rotational frequency at which the star remains non chemically homogeneous (upward triangles), and the minimum rotational frequency at which the star becomes chemically homogeneous (downward triangles), for a grid of masses ranging from 10 to 150 $M_\odot$ and three metallicities, $Z=0.01,\ 0.001,\ 0.0001$.

The following fits for the threshold angular frequency for chemically homogeneous evolution are implemented in \COMPAS and shown in Figure~\ref{fig:CHEfits}:

\setlength{\abovedisplayskip}{1pt}
\Large
\begin{equation}
    \omega_{\text{M,Z}}=\frac{\omega_{\text{M},{\text{Z$_{0.004}$}}}}{\scriptstyle0.09\ln\textstyle(\frac{\text{Z}}{0.004})\scriptstyle+1}
\label{eq:CHEthreshold1}
\end{equation}

\normalsize\noindent
where
\Large
\begin{equation}
    \omega_{\text{M},{\text{Z$_{0.004}$}}}=
    \begin{cases}    
        \sum_{i=0}^5 a_i \frac{\text{M}^i}{\text{M}^{0.4}} &\hspace{-0.1cm}\scriptstyle\text{\text{rad}\ $\text{s}^{\minus{1}}$},\text{M}\leq 100\,\text{M}_\odot\\
        \sum_{i=0}^5 a_i \frac{100^i}{\text{M}^{0.4}} &\hspace{-0.1cm}\scriptstyle\text{\text{rad}\ $\text{s}^{\minus{1}}$},\text{M}>100 \text{M}_\odot\\
    \end{cases}
\label{eq:CHEthreshold2}
\end{equation}

\normalsize\noindent
and

\begin{itemize}[label={}, labelwidth =\widthof{\bfseries9999}, leftmargin = !, noitemsep, topsep = 0pt]
    \item$a_0$=~5.7914\tenpow{\minus4}
    \item$a_1$=\minus1.9196\tenpow{\minus6}
    \item$a_2$=\minus4.0602\tenpow{\minus7}
    \item$a_3$=~1.0150e\tenpow{\minus8}
    \item$a_4$=\minus9.1792\tenpow{\minus11}
    \item$a_5$=~2.9051\tenpow{\minus13}
\end{itemize}

We expect these fits to be valid over the range where they are constructed ($10\ M_\odot \leq M \leq 150 M_\odot$, $10^{-4} \leq Z \leq 0.01$) but caution should be exercised if the fits are extrapolated significantly beyond these boundaries. 

\begin{figure}
    \centering
	\includegraphics[width=8.0cm]{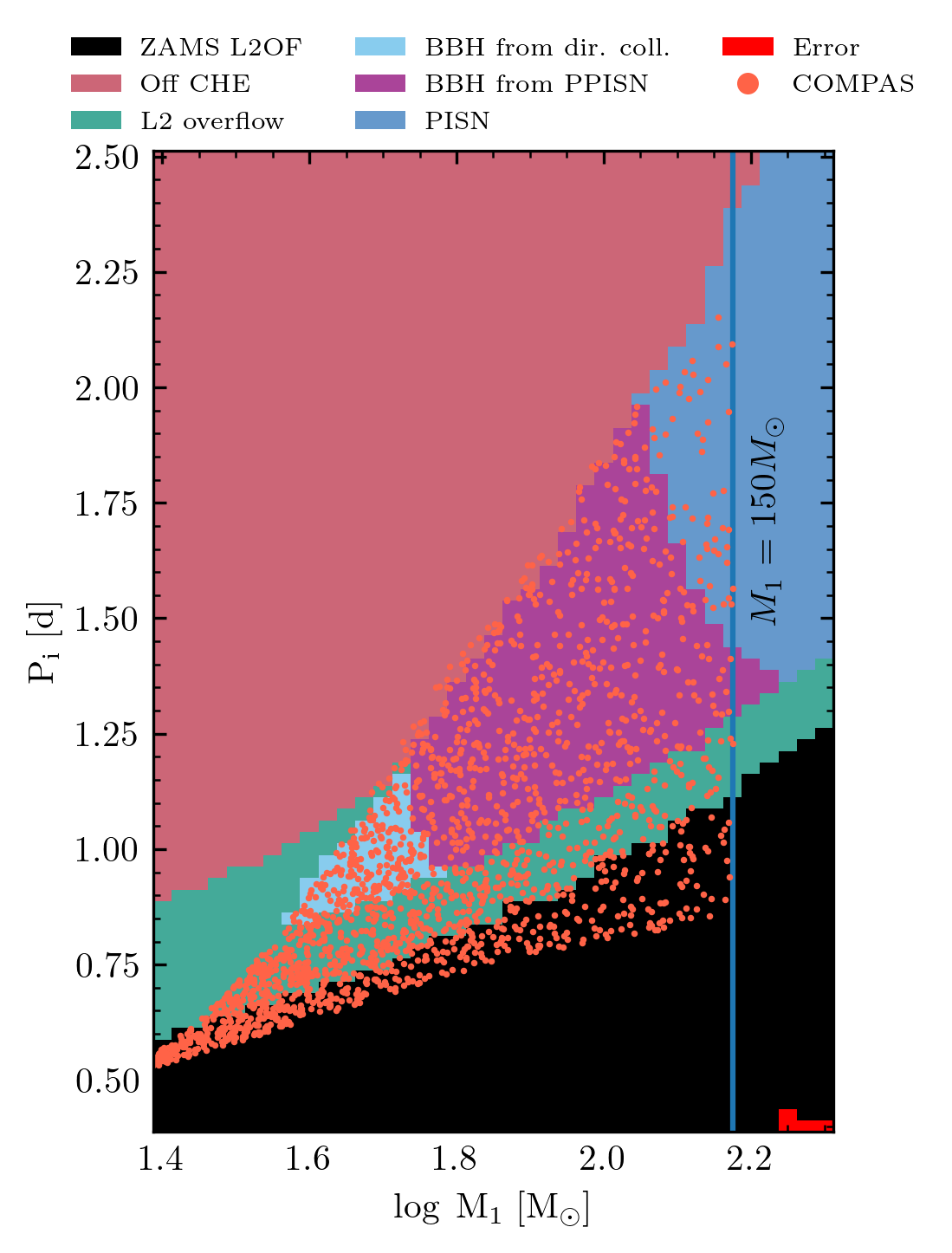}
	\vspace{-11pt}
	\captionof{figure}{Comparison between the \COMPAS mass and orbital period range leading to \ac{CHE}, implemented as described in this appendix (red dots), and the detailed MESA binary models of \citet{duBuisson_2020} (background colour shading of cyan, blue and purple indicates binaries that survive \ac{CHE} on the main sequence) at $Z=0.001$.}
	\label{fig:CHEcomparison}
\end{figure}

Figure~\ref{fig:CHEcomparison} shows the range of binaries in mass -- orbital period space that lead to \ac{CHE} at $Z=0.001$.  Red points indicate the population of binaries that remain \ac{CH} through the main sequence according to the model described here and implemented in \ac{COMPAS}.  For comparison, the background colours show the outcomes from detailed MESA binary models of \citet{duBuisson_2020}, with cyan, purple and blue denoting binaries that undergo \ac{CHE} on the main sequence without merging.  In general, there is good agreement between the two sets of models, particularly at higher orbital periods.  For the tightest orbits, \citet{duBuisson_2020} models predict mergers through L2 overflow when our \ac{COMPAS} fits suggest that the over-contact binary may still survive with \ac{CHE} stars.   This is likely due to the combination of rotational and tidal deformation, as well as some mild expansion early during \ac{CHE} evolution, that are not accounted for in the \ac{COMPAS} models.

\label{lastpage}
\end{document}